\documentclass[floats,floatfix,showpacs,amssymb,prd,twocolumn,superscriptaddress,nofootinbib,longbibliography]{revtex4-1}

\usepackage{anyfontsize}
\usepackage{amssymb}
\usepackage{amsmath}
\usepackage{amsfonts}
\usepackage{amsbsy}
\usepackage[utf8]{inputenc} 
\usepackage{newtxtext}
\usepackage{newtxmath}

\usepackage{tensor}
\usepackage{mathrsfs}

\usepackage{bm}

\usepackage[utf8]{inputenc}

\usepackage{graphicx}
\usepackage{epsfig}
\usepackage{epstopdf}

\usepackage[usenames,dvipsnames]{xcolor}

\usepackage{verbatim,mathtools,needspace,enumitem,etoolbox}

\definecolor{linkcolor}{rgb}{0.0,0.3,0.5}

\usepackage[unicode, colorlinks=true, linkcolor=linkcolor, citecolor=linkcolor, filecolor=linkcolor,urlcolor=linkcolor, pdfusetitle]{hyperref}

\usepackage{epstopdf}

\usepackage{bm}
\usepackage{dcolumn}

\usepackage{longtable}
\usepackage{enumerate}
 \usepackage[normalem]{ulem}

\hypersetup{colorlinks=true,
            citecolor=blue,
            linkcolor=blue,
            urlcolor=blue}
\usepackage[T1]{fontenc}
\usepackage{subfigure}
\usepackage{amsmath}
\usepackage{float}
\usepackage{mathrsfs}
\usepackage[dvipsnames]{xcolor}
\usepackage{soul}

\newcommand{\f}{\left(1-\frac{2\,M}{\sqrt{r^2+a^2}}\right)}
\definecolor{darkgreen}{RGB}{1,212,57}

\setcitestyle{numbers,square}

\setcitestyle{numbers,square}

\begin{document}
\title{Scalar absorption: Black holes versus wormholes
}

\author{Haroldo C. D. Lima Junior}
\email{haroldo.ufpa@gmail.com} 
\affiliation{Faculdade de F\'{\i}sica,
Universidade Federal do Par\'a, 66075-110, Bel\'em, PA, Brazil }

\author{Carolina L. Benone}%
 \email{benone@ufpa.br}
\affiliation{%
 Campus Universit{\'a}rio Salin{\'o}polis, Universidade Federal do Par{\'a}, 68721-000, Salin{\'o}polis, Par{\'a}, Brazil 
}%

\author{Lu\'{\i}s C. B. Crispino}
\email{crispino@ufpa.br} 
\affiliation{Faculdade de F\'{\i}sica,
Universidade Federal do Par\'a, 66075-110, Bel\'em, PA, Brazil }

\date{\today}

\begin{abstract}
We study the absorption of massless scalar waves in a geometry that interpolates between the Schwarzschild solution and a wormhole that belongs to the Morris-Thorne class of solutions. In the middle of the interpolation branch, this geometry describes a regular black hole. We use the partial wave approach to compute the scalar absorption cross section in this geometry. Our results show that black holes and wormholes present distinctive absorption spectra. We conclude, for instance, that the wormhole results are characterized by the existence of quasibound states which generate Breit-Wigner-like resonances in the absorption spectrum. 
\end{abstract}

\maketitle

\section{Introduction}
Recently, an increasing number of observations concerning  gravity in the strong field regime has been presented in the literature. The LIGO/VIRGO collaborations have cataloged many gravitational waves (GWs) detections, mostly associated to binary black hole (BH) mergers (see, for instance, \cite{VL1,VL2,VL3,VL4}). On the other hand, the international collaboration Event Horizon Telescope (EHT) presented the first ever image of the shadow of a supermassive BH \cite{EHT}. These observations of GWs and the shadow of a BH confirmed the results predicted by the theory of General Relativity (GR).

Within GR, BHs are solutions of the Einstein's field equations that posses an event horizon. The first exact solution of Einstein's equation is known as Schwarzschild geometry~\cite{Sch}, which describes a spherically symmetric, electrically uncharged and non-rotating BH. The spherically symmetric, electrically charged and non-rotating BH geometry is known as the Reissner-Nordstr\"om solution~\cite{Reissner,Nordstrom}. The first exact uncharged rotating BH solution was obtained by Kerr \cite{Kerr:1963}, while the charged rotating BH was presented in Ref.~\cite{kerrnewman}. Such {\it standard} BH solutions of GR are cursed with singularities, where geometrical quantities diverge and physics predictability breaks down. Under some conditions, singularities are expected to exist within GR, as stated by the so called singularity theorems \cite{Penrose:1964wq,Singularitytheo}.

Regular BHs, i.e. non-singular BH solutions, were proposed as an alternative to avoid the singularity problem. The geometric quantities associated to such solutions are finite everywhere. The first regular BH solution was proposed by Bardeen \cite{bardeen}. After that, many others regular BH solutions were presented (see, for instance, Refs.~\cite{Hayward,ayonb,RBH}). 

Wormholes are solutions that connect two asymptotically flat  regions by a throat \cite{Matt_Visser}. In fact, the Schwarzschild solution itself can be associated to a wormhole, known as Einstein-Rosen bridge, which is not traversable~\cite{Einstein_Rosen,dinverno}. A well-known class of wormhole solutions were proposed by Morris and Thorne \cite{MorrisThorne}, and they can be, in principle, traversable. The traversability of the Morris-Thorne class of solutions can be understood as a consequence of exotic matter violating the weak energy condition at the wormhole throat \cite{MorrisThorne,Morris-Thorne-Yurtsever,Matt_Visser::1989}. 

Absorption of matter and fields is of great interest in GR, for instance, in explaining the role of accretion by BHs in active galactic nuclei \cite{AGN,AGN1,AGN2}. Besides that, the results for the absorption cross section in the high-frequency limit are related to the shadows of BHs. The absorption of scalar waves by a Schwarzschild BH was studied in Ref.~\cite{Abs1}. Similar studies for Reissner-Nordstr\"om \cite{Abs2,Abs4}, Bardeen  \cite{Abs5}, Kerr \cite{Abs6}, and Kerr-Newman \cite{Abs8,Abs9,Abs10} BHs can also be found in the literature. In comparison to  BHs, few results for the absorption of scalar waves by wormhole spacetimes are available (see, for instance, Refs.~\cite{Abs11,PRD:100024016:2019}).
 
We investigate the propagation of planar massless scalar waves in the geometry proposed by Simpson and Visser in Ref.~\cite{Black_bounce}. The line element of this geometry depends on two parameters. Depending on the values of such parameters, this geometry describes a Schwarzschild BH, a regular BH, or a wormhole spacetime belonging to the Morris-Thorne class. In the regular BH branch of interpolation, instead of a singularity there is a spacelike hypersurface, which represents a bounce into a future incarnation of the universe. The quasinormal modes of this solution were recently analyzed in Ref.~\cite{QNM_SV} and similar {\it black-bounce} solutions were proposed in Ref.~\cite{Black_bounce2}. We study the scalar absorption for Schwarzschild and regular BHs, as well as for Morris-Thorne wormholes, considering the Simpson-Visser line element.

The absorption spectrum of this geometry presents interesting features. For instance, the wormhole solution can show imprints of quasibound states around the throat, leading to narrow peaks in the absorption spectrum. 
Similar results appear for extreme/exotic compact objects (ECOs)~\cite{PRD:98104034:2018} and BH remnants in the context of metric-affine gravity~\cite{PRD:100024016:2019}.

The remainder of this paper is organized as follows: In Sec.~\ref{Sec1}, we review the main properties of the Simpson-Visser geometry. In Sec.~\ref{Sec2}, we outline the partial wave method, valid for both the BH and the wormhole cases. In Sec.~\ref{HFR}, we investigate the high-frequency regime. In Sec.~\ref{LHF}, we present the results for the absorption of  massless scalar waves by the BH branch of interpolation, while the results for the wormhole case are presented in Sec.~\ref{Sec5}. Our final remarks are presented in Sec.~\ref{Sec6}. We use natural units, such that $G=c=\hbar=1$, and we adopt the metric signature ($-$,$\ +$,$\ +$,$\ +$).

\section{Spacetime description}
\label{Sec1}
\hspace{0.4cm}The Simpson-Visser spacetime is described by the following line element \cite{Black_bounce}: 
\begin{align}
\nonumber ds^2=&-\left(1-\frac{2\,M}{\sqrt{r^2+a^2}}\right)\,dt^2+\left(1-\frac{2\,M}{\sqrt{r^2+a^2}}\right)^{-1}\,dr^2\\
\label{Lineel}&+\left(r^2+a^2\right)\,\left(d\theta^2+\sin^2\theta\,d\varphi^2\right).
\end{align}
Depending on the value of the parameter $a\geq 0$, we have \cite{Black_bounce}:
\begin{itemize}
\item A Schwarzschild BH spacetime ($a=0$);
\item A regular BH spacetime ($0<a<2\,M$);
\item A one-way traversable wormhole geometry with a null throat ($a=2\,M$);
\item A two-way traversable wormhole geometry belonging to the Morris-Thorne class ($a > 2\,M$).
\end{itemize}

For the regular BH branch of interpolation, the BH event horizon is located where $g^{r\,r}=0$, corresponding to
\begin{equation}
\label{EH}r_h=\sqrt{(2\,M)^2-a^2}.
\end{equation}
\\ 
As pointed out in Ref.~\cite{Black_bounce}, the curvature invariants of the Simpson-Visser geometry are regular everywhere for $a\neq 0$.
We can analyze massless scalar waves propagating in this Simpson-Visser geometry, using an unified description, which can be applied to a BH or a wormhole by choosing the appropriate values for the parameter $a$. 

\section{Partial wave analysis}
\label{Sec2}

\begin{center}
\begin{figure}[ht!]
\center
\includegraphics[scale=0.6]{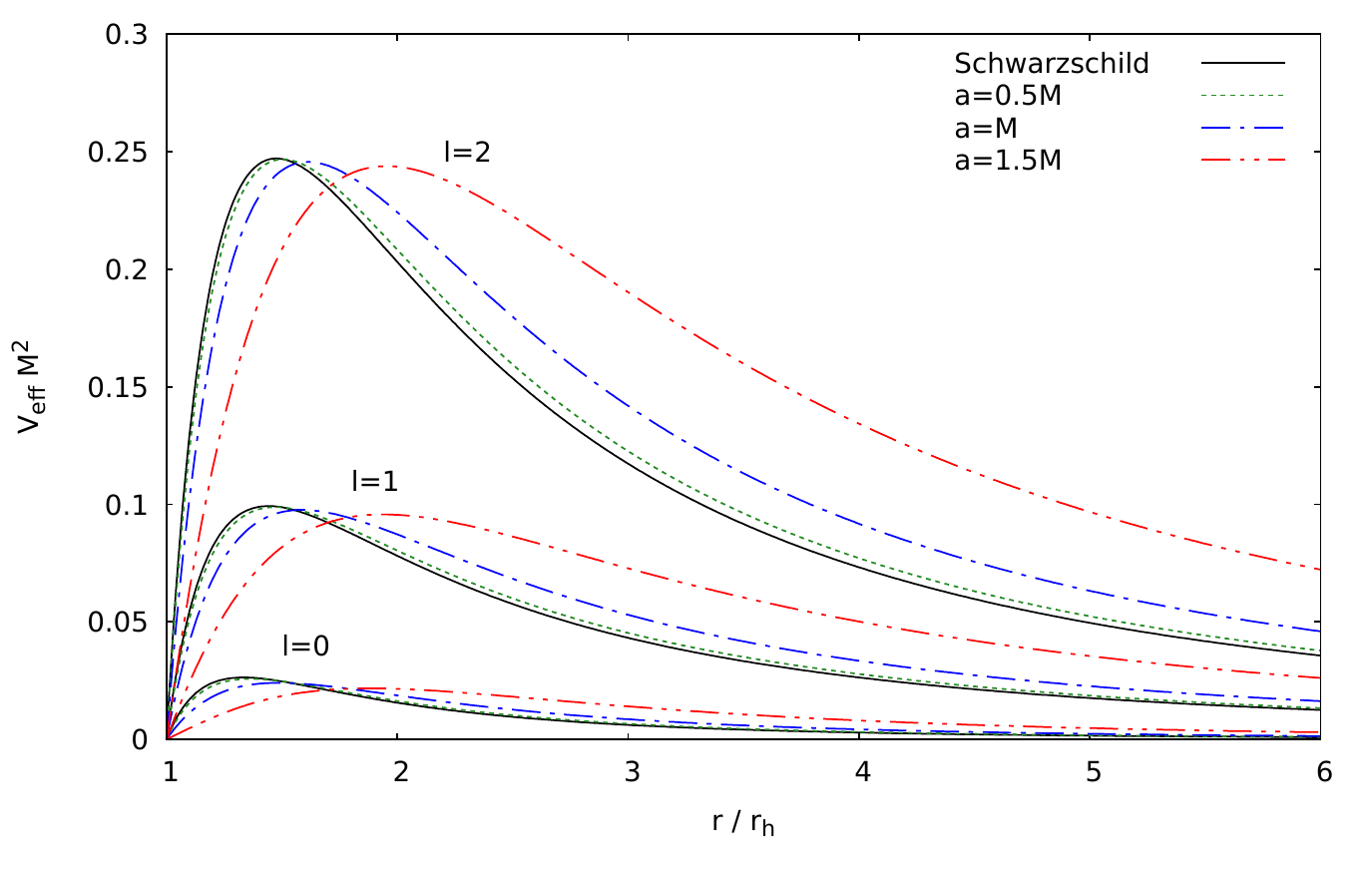}
\caption{Effective potential, given by Eq.~\eqref{veff}, for the massless scalar field $\phi$ in the BH case, as a function of the radial coordinate $r$ in units of the event horizon $r_h$, described in Eq.~\eqref{EH}. In this figure, we have selected different values for the parameter $a$, obeying $0 \leq a < 2\,M$.}
\label{Veff}
\end{figure}
\end{center}

\begin{center}
\begin{figure}[ht!]
\center
\includegraphics[scale=0.55]{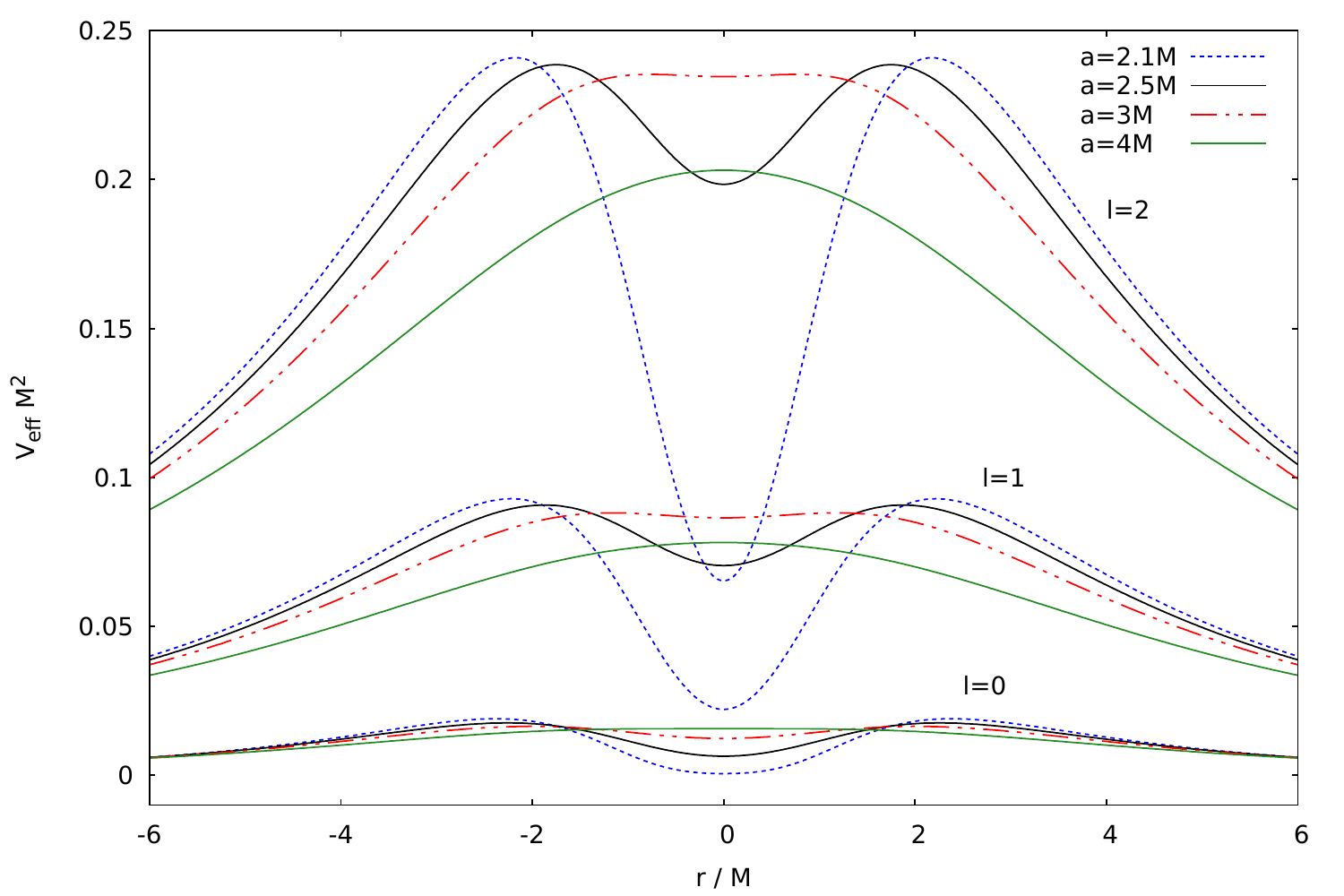}
\caption{Effective potential, given by Eq.~\eqref{veff}, for the massless scalar field $\phi$ in the wormhole case, as a function of the radial coordinate $r$ in units of $M$. In this figure, we have selected different values for the parameter $a$, obeying $a\geq2\,M$. 
}
\label{Veff2}
\end{figure}
\end{center}

In a curved spacetime, the Lagrangian for the minimally coupled massless scalar field $\Phi$ is
\begin{equation}
\mathcal{L}=\frac{1}{2}\nabla_\mu\Phi\,\nabla^\mu\Phi.
\end{equation}
The equations of motion for the massless scalar field are obtained through the Euler-Lagrange procedure, resulting in
\begin{equation}
\nabla^\mu\,\nabla_\mu\,\Phi=0, 
\end{equation}
which we may rewrite as
\begin{equation}
\label{eqmotionfield}\frac{1}{\sqrt{\left|g\right|}}\partial_\mu\left(\sqrt{\left|g\right|}\,g^{\mu\nu}\partial_\nu\Phi\right)=0,
\end{equation}
where $g^{\mu\nu}$ are the contravariant components of the metric tensor and $g$ is the metric determinant. A monochromatic scalar wave with frequency $\omega$ propagating in the spacetime described by Eq.~\eqref{Lineel} may be written as
\begin{equation}
\label{modes}\Phi=\sum_{l,m}\frac{\phi(r)}{\left(r^2+a^2\right)^\frac{1}{2}}\,Y_{ l m}(\theta,\varphi)\,e^{-i\,\omega\,t},
\end{equation}
where $Y_{ l m}(\theta,\varphi)$ are the scalar spherical harmonics.
Substituting Eq.~\eqref{modes} in Eq.~\eqref{eqmotionfield}, and using the well-known properties of the scalar spherical harmonics, we find an ordinary differential equation for $\phi(r)$, namely
\begin{equation}
\label{req}f(r)\,\frac{d}{d\,r}\left(f(r)\,\frac{d\phi(r)}{d\,r}\right)+\left[\omega^2-V_{\textit{eff}}\right]\,\phi(r)=0,
\end{equation}
where 
\begin{equation}
f(r)\equiv\f,
\end{equation}
and $V_{\textit{eff}}$ is the effective scattering potential for the massless scalar waves, given by
\begin{align}
\label{veff} V_{\textit{eff}}\equiv f(r)\,\left[\frac{f'(r)\,r}{\left(r^2+a^2\right)}+\frac{a^2\,f(r)}{\left(r^2+a^2\right)^2}+\frac{l\,(l+1)}{\left(r^2+a^2\right)}\right].
\end{align}
Here the prime denotes differentiation with respect to the radial coordinate $r$. Plots of the effective potential $V_{\textit{eff}}$ for different choices of $a$ are shown in Figs.~\ref{Veff} and \ref{Veff2}. In Fig.~\ref{Veff} we show the effective potential for the BH branch of interpolation, while in Fig.~\ref{Veff2} we show the effective potential for the wormhole case.

In order to study the solutions of Eq.~\eqref{req}, we define the tortoise coordinate $x$, given by
\begin{equation}
dx\equiv f^{-1}(r)\,dr.
\end{equation}
The radial equation \eqref{req}, in terms of the tortoise coordinate, can be rewritten as
\begin{equation}
\label{rad_eq}\left(\frac{d^2}{dx^2}+\omega^2-V_{\textit{eff}}\right)\,\phi(x)=0,
\end{equation}
which is a Schr{\"o}dinger-like equation subjected to the radial scattering potential $V_{\textit{eff}}$. In the next sections we treat in details the absorption of massless scalar waves by the BH and the wormhole cases, both described by Eq.~\eqref{rad_eq}.

\begin{center}
\begin{figure}[ht!]
\center
\includegraphics[scale=0.34]{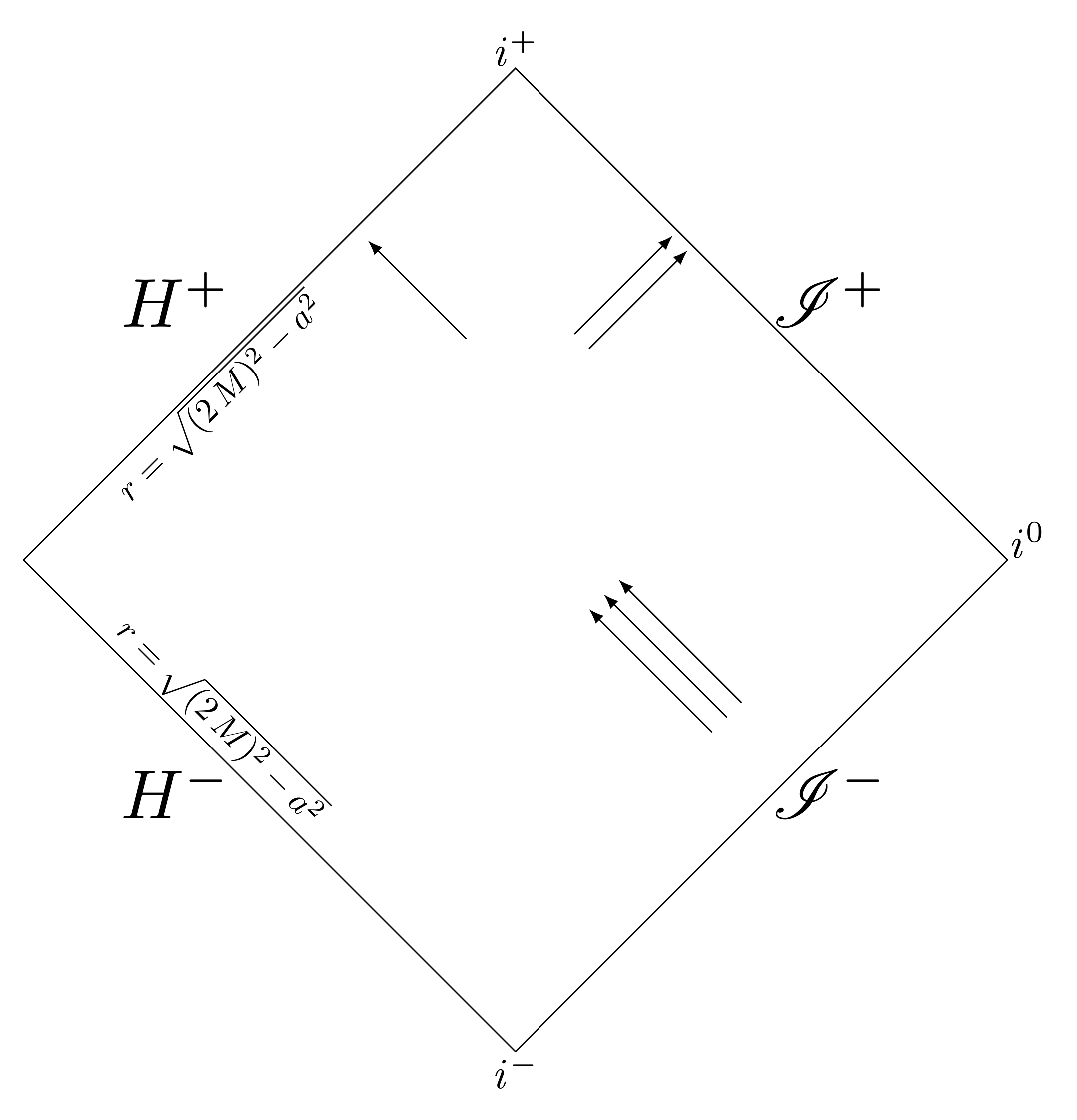}
\caption{Carter-Penrose diagram for the BH branch of interpolation ($0\leq a< 2\,M$).  The arrows represent the scattering problem of the massless scalar waves, being partly absorbed by the BH and partly reflected to the future null infinity.}
\label{CP_Diag_BH}
\end{figure}
\end{center}

In order to solve Eq.~\eqref{rad_eq}, we need to impose boundary conditions. Such boundary conditions must be consistent with the physical problem that we are handling. For the BH branch of interpolation, we are interested in solutions representing a scalar wave incoming from the past null infinity ($\mathscr{I}^-$). Moreover, we assume that there are only ingoing scalar waves at the future event horizon ($H^+$), since it is a one-way membrane. The waves reflected by the scattering potential propagate to the future null infinity ($\mathscr{I}^+$). This may be expressed in terms of the \textit{in modes}, described by

\begin{equation}
\label{BC_BH}\phi(r) \approx \begin{cases} e^{-i \omega x}+e^{i\omega x}R_{\omega l}, & r \rightarrow +\infty \ \ (x\rightarrow +\infty),\\
T_{\omega l}\,e^{-i\omega x}, & r \rightarrow r_h\ \ \ \ (x \rightarrow -\infty).
\end{cases}
\end{equation}
The Carter-Penrose diagram, outside the event horizon of the BH, with the illustration of this BH scattering problem, is presented in Fig.~\ref{CP_Diag_BH}. The reflection and transmission coefficients are given by $\left|R_{\omega l}\right|^2$ and $\left|T_{\omega l}\right|^2$, respectively. 

\begin{center}
\begin{figure}[ht!]
\center
\includegraphics[scale=0.34]{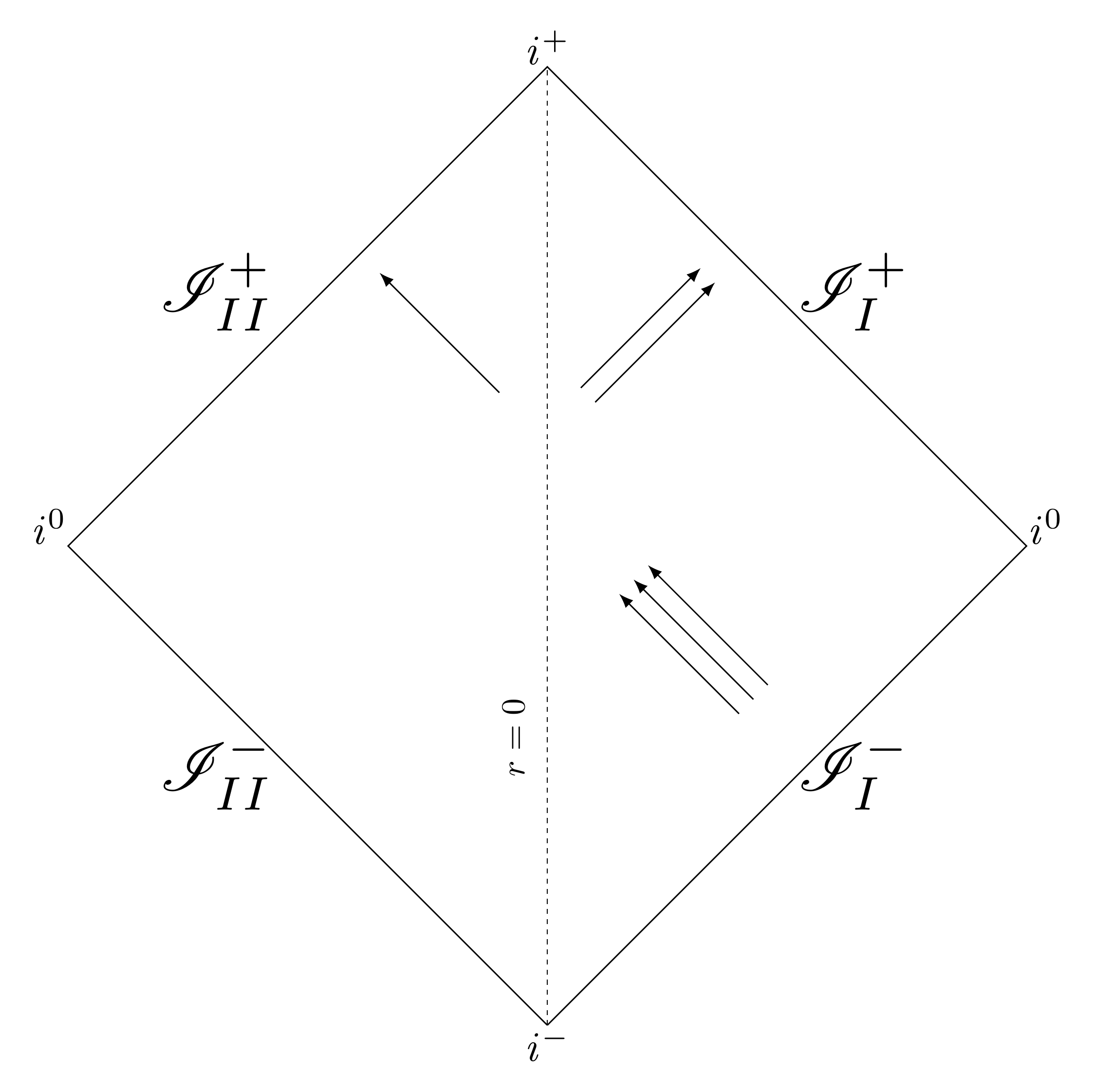}
\caption{Carter-Penrose diagram for the wormhole branch of interpolation ($a\geq 2\,M$). The arrows represent the scattering problem of the massless scalar waves, being partly transmitted through the wormhole's throat, located at $r=0$,  and partly reflected to the future null infinity. }
\label{CP_Diag_wmhl}
\end{figure}
\end{center}

For the wormhole branch of interpolation, there is no event horizon. In this case, we are interested in solutions representing a scalar wave incoming from the past null infinity $\mathscr{I}_I^-$, being transmitted through the wormhole's throat to the future null infinity $\mathscr{I}_{II}^+$, or reflected to the future null infinity $\mathscr{I}_I^+$. The Carter-Penrose diagram with the illustration of this wormhole scattering problem is represented in Fig.~\ref{CP_Diag_wmhl}. Such boundary conditions may be expressed as:
\begin{equation}
\phi(r) \approx \begin{cases} e^{-i \omega x}+R_{\omega l}\,e^{i\omega x}, & r \rightarrow +\infty \ \ (x\rightarrow +\infty),\\
T_{\omega l}\,e^{-i \omega x}, & r \rightarrow -\infty \ \ \ (x \rightarrow -\infty).
\end{cases}
\end{equation}

Using the partial wave method \cite{Futterman}, the total absorption cross section for the BH, as well as for the wormhole, can be written as
\begin{equation}
\label{tot_abs}\sigma_{abs}=\sum_{l=0}^{\infty}\,\sigma_l,
\end{equation}
where
\begin{equation}
\label{partial_abs}\sigma_l=\frac{\pi}{\omega^2}\,\left(2\,l+1\right)\,\left|T_{\omega\,l}\right|^2
\end{equation}
is the partial absorption cross section, i.e., the absorption cross section for a fixed $l$.

\section{High-frequency regime}
\label{HFR}
In the high-frequency regime, the massless scalar field can be described by the null geodesic equation. Within this regime, the absorption cross section tends to the classical capture cross section of null geodesics. 

\begin{center}
\begin{figure}[ht!]
\center
\includegraphics[scale=0.6]{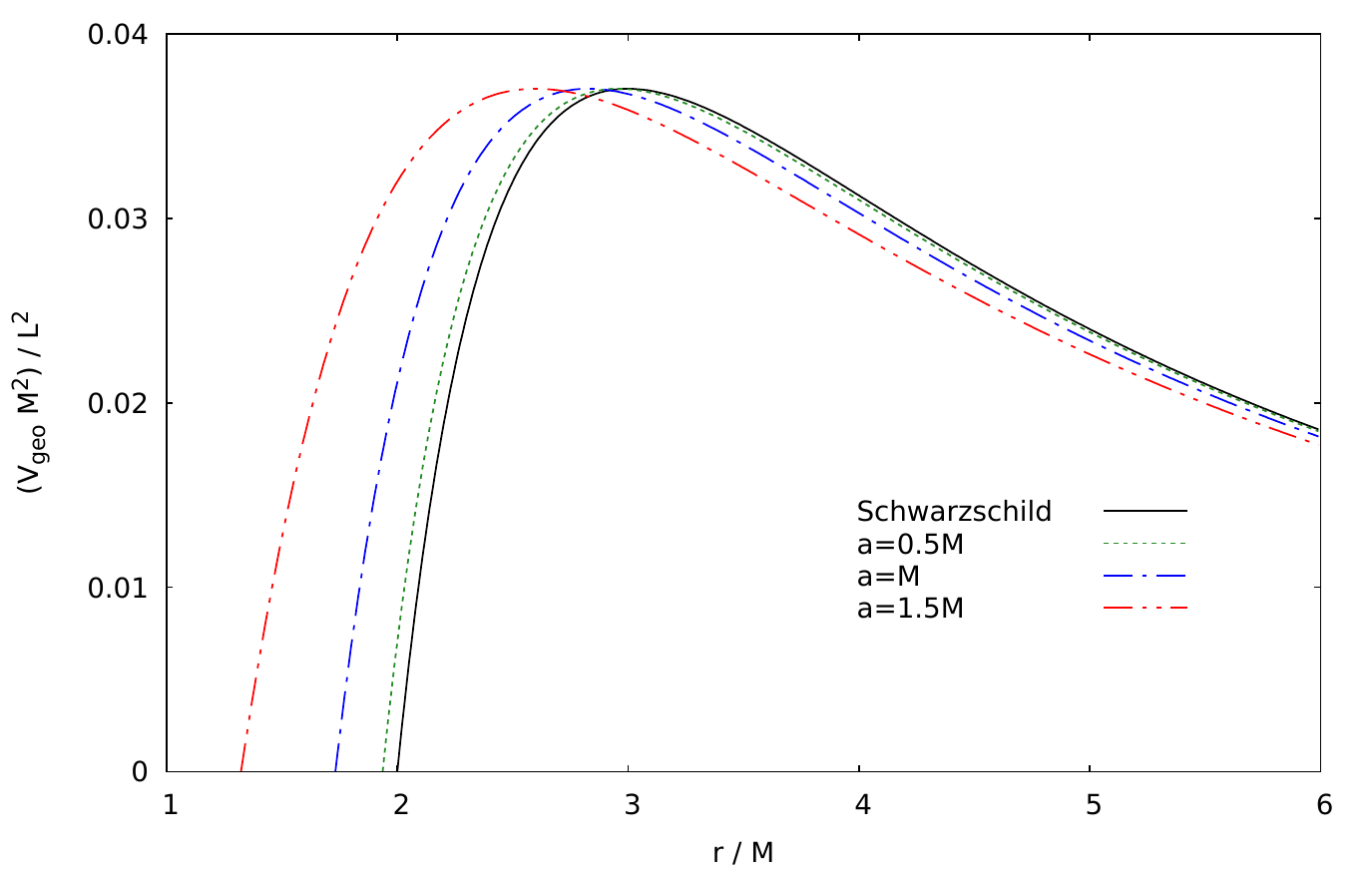}
\caption{Effective potential, given by Eq.~\eqref{Veffgeo},  for null geodesics in the BH case, as a function of the radial coordinate $r$. In this figure, we have selected different values for the parameter $a$ ($0 \leq a < 2M$).}
\label{VeffBH_geo}
\end{figure}
\end{center}

\begin{center}
\begin{figure}[ht!]
\center
\includegraphics[scale=0.6]{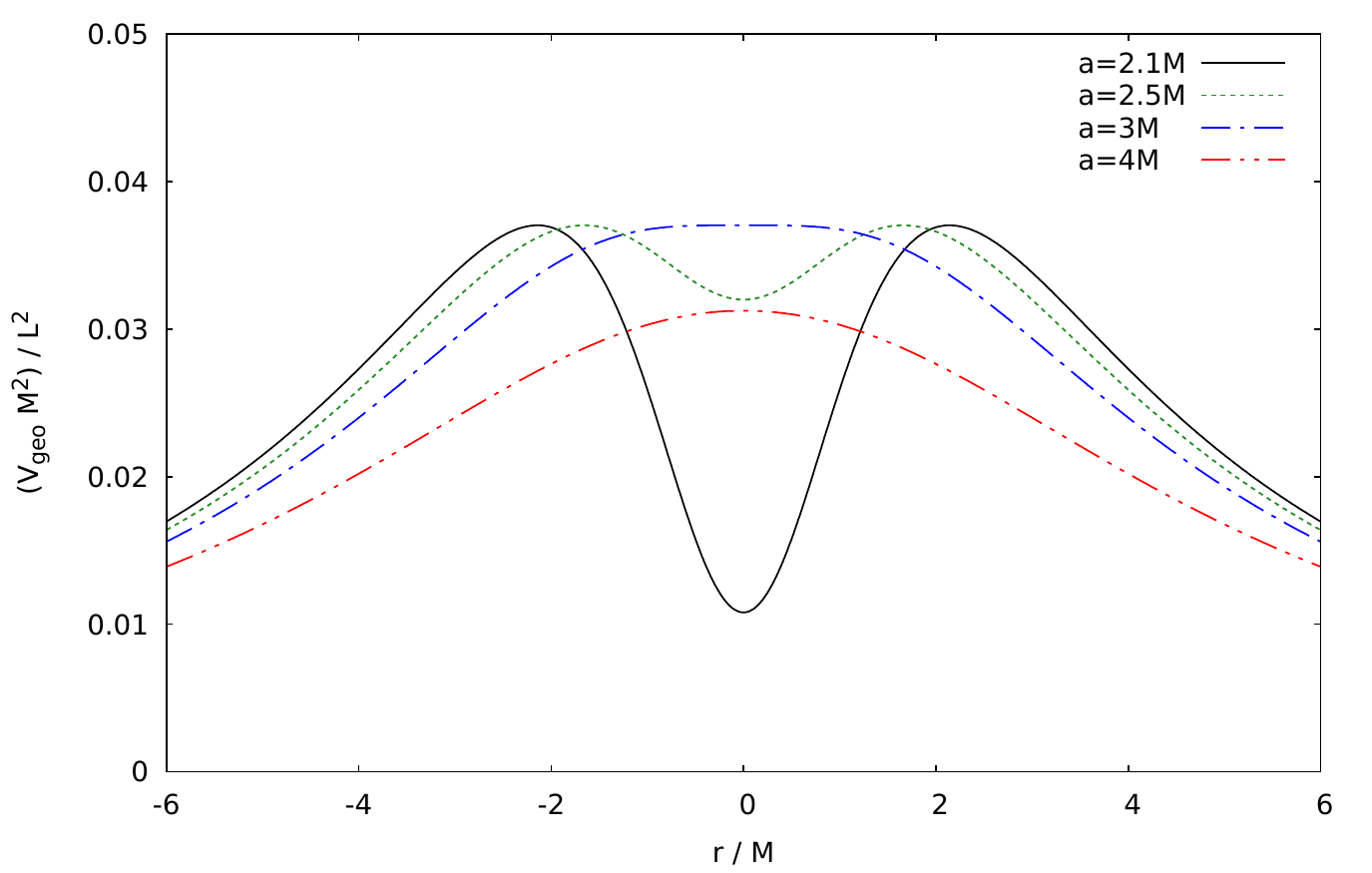}
\caption{Effective potential, given by Eq.~\eqref{Veffgeo}, for null geodesics in the wormhole case, as a function of the radial coordinate $r$.  In this figure, we have selected different values for the parameter $a$ ($a\geq2M$).}
\label{Veffwhml_geo}
\end{figure}
\end{center}

Here, we summarize the main results of the propagation of null geodesics in the spacetime described by the line element \eqref{Lineel}. We consider the Lagrangian $\mathcal{L}_{geo}$, that obeys
\begin{align}
&2\,\mathcal{L}_{geo}=g_{\mu\nu}\,\dot{x}^\mu\,\dot{x}^\nu\therefore\\
\label{Lgeo}&2\,\mathcal{L}_{geo}=-f(r)\,\dot{t}^2+f^{-1}(r)\,\dot{r}^2+(r^2+a^2)\,\dot{\varphi}^2=0,
\end{align}
where we set, without loss of generality, $\theta=\pi/2$. The overdots represent derivatives with respect to the affine parameter along the null geodesics. Since the Lagrangian $\mathcal{L}_{geo}$ does not depend on $t$ and $\varphi$, we have two conserved quantities, namely
\begin{align}
\label{Energy}&E=\f\,\dot{t},\\
\label{Angmomentum}&L=\left(r^2+a^2\right)\,\dot{\varphi}.
\end{align}
$E$ and $L$ are the energy and angular momentum, respectively, measured by an observer at infinity. By substituting Eqs.~\eqref{Energy} and \eqref{Angmomentum} in Eq.~\eqref{Lgeo}, we find 
\begin{equation}
\label{rdot}\dot{r}^2+V_{geo}=E^2,
\end{equation}
which is an energy balance equation, with $V_{geo}$ being the effective potential for null geodesics, given by
\begin{equation}
\label{Veffgeo}V_{geo}\equiv\f\,\frac{L^2}{(r^2+a^2)}.
\end{equation}
In Figs.~\ref{VeffBH_geo} and \ref{Veffwhml_geo}, we show the effective potential for null geodesics, for the BH and the wormhole cases, respectively. In Fig.~\ref{Veffwhml_geo}, we note the existence of a minimum in the effective potential $V_{geo}$, which is associated to stable null geodesics at $r=0$.

In the high-frequency limit, the absorption cross section of the massless scalar field tends to the capture cross section, also called geometric cross section ($\sigma_{geo}$), which is given by \cite{Wald}
\begin{equation}
\label{siggeo}\sigma_{geo}=\pi\,b_c^2,
\end{equation}
where $\left. b_c\equiv L/E\right|_{r_{ph}}$ is the critical impact parameter, computed through
\begin{align}
\label{photonorbit1}&\left. V_{geo}\right|_{r=r_{ph}}=E^2,\\
\label{photonorbit2}&\left.\frac{d V_{geo}}{dr}\right|_{r=r_{ph}}=0.
\end{align}
The value $r=r_{ph}$ corresponds to the location of unstable circular photon orbits in the Simpson-Visser spacetime. The solutions  of Eqs.~\eqref{photonorbit1}-\eqref{photonorbit2} are 
\begin{align}
\label{LR}r_{ph}=\begin{cases}
 \sqrt{9\,M^2-a^2}, & \text{if\ \ } 0\leq a< 2\,M,\\
\pm \sqrt{9\,M^2-a^2}, & \text{if\ \ } 2\,M\leq a\leq 3\,M,\\
0, & \text{if\ \ } a> 3\,M.
\end{cases}
\end{align}

Therefore, we may have two different critical impact parameters $b_0$ and $b_1$ associated to unstable circular photon orbits, depending on the value of the parameter $a$. The expressions for $b_0$ and $b_1$ are given by:
\begin{align}
\label{bc0}&b_0=\left. \frac{L}{E}\right|_{r=0}=\frac{a^{\frac{3}{2}}}{\left(a-2\,M\right)^\frac{1}{2}},\\
\label{bc1}&b_1=\left. \frac{L}{E}\right|_{r=\sqrt{9\,M^2-a^2}}=3\sqrt{3}\,M.
\end{align}
Thus, the geometric cross section is given by:
\begin{equation}
\label{Geo_Cross_BH}\sigma_{geo}=\begin{cases}\pi\,b_1^2=27\,\pi\,M^2, & \text{if\ \ } 0\leq a \leq 3\,M\\
\pi\,b_0^2=\pi\,\left(\frac{a^3}{a-2\,M}\right), & \text{if\ \ }  a > 3\,M.
\end{cases}
\end{equation}
We point out that for $0\leq a \leq 3\,M$, the geometric cross section is independent of $a$ and coincides with the Schwarzschild result. On the other hand if $a>3\,M$, the geometric cross section depends on $a$.
\begin{center}
\begin{figure}[ht!]
\center
\includegraphics[scale=0.68]{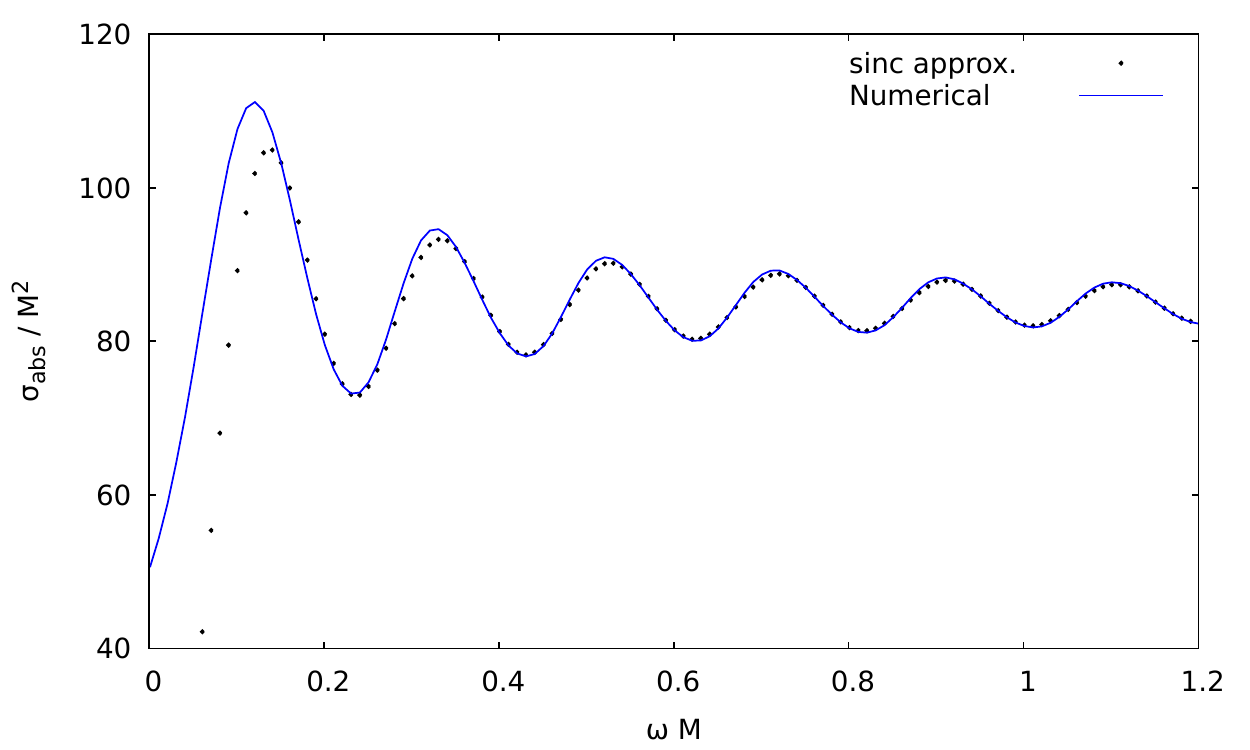}
\caption{Comparison between the numerical result for the total absorption cross section of a regular BH described by the Simpson-Visser line element and the sinc approximation [Eq.~\eqref{tot_abs_sinc}]. In this figure, we have chosen $a=0.5M$.}
\label{sinc_apprx}
\end{figure}
\end{center}

\section{Black hole results}
\label{LHF}

\subsection{Low-frequency regime}
\label{Lowf_regime}
It is known that in the low-frequency regime the absorption cross section of massless scalar fields for spherically symmetric BHs tends to the area of the BH horizon \cite{Das_Gibbons_Mathur}. In fact, this result is also true for stationary BHs \cite{AH1}. The area of the BH horizon associated to the line element \eqref{Lineel} is given by
\begin{align}
&A_h=\left.\int_0^{\pi}\,\int_{-\pi}^{\pi}\sqrt{g_{\theta \theta}\,g_{\varphi\varphi}}d\varphi\,d\theta\right|_{r=r_h}\therefore\\
\label{EH_area}&A_h=\left.4\,\pi\left(r^2+a^2\right)\right|_{r=r_h}=16\,\pi\,M^2.
\end{align}
We conclude that the area of the regular Simpson-Visser BH horizon is independent of $a$ and equal to the Schwarzschild one. Therefore, in the low-frequency regime, we expect the total BH absorption cross section to have the same limit regardless of the parameter $a$.

\subsection{Sinc approximation}

We can improve the high-frequency regime analysis for BHs by taking into consideration the sinc approximation. As shown in Ref.~\cite{Sinc_approx}, following the result originally obtained in Ref.~\cite{Abs1}, the oscillatory part of the absorption cross section is given by
\begin{equation}
\sigma_{osc}=-\frac{8\,\pi\,\lambda_l}{\Omega_l}\,e^{-\frac{\pi\,\lambda_l}{\Omega_l}}\,\text{sinc}\left(\frac{2\,\pi\,\omega}{\Omega_l}\right)\,\sigma_{geo},
\end{equation}
where $\text{sinc}(x)\equiv\sin{(x)}/x$,
\begin{equation}
\lambda_l = \frac{1}{\dot{t}}\sqrt{\frac{1}{2}\frac{d^2 (\dot{r}^2)}{dr^2}},
\end{equation}
is the Lyapunov exponent of the null geodesic \cite{Sinc_approx,Sinc_approx2}, and
\begin{equation} 
\label{sinc} \Omega_l=\frac{d\,\varphi}{dt}=\sqrt{\frac{f(r_{ph})}{r_{ph}^2+a^2}},
\end{equation}
is the coordinate angular velocity of the null geodesic.

The absorption cross section in the high-frequency regime can be approximated by \cite{Sinc_approx}:
\begin{equation}
\label{tot_abs_sinc}\sigma_{hf}\approx \sigma_{geo}+\sigma_{osc}.
\end{equation}

In Fig.~\ref{sinc_apprx}, we present a comparison between the  numerical result, obtained in Sec.~\ref{Num_res}, and the sinc approximation. The results agree remarkably well for high frequencies, being very close to each other even for intermediate values of the frequency. 

\subsection{Numerical results}
\label{Num_res}

Let us now present the results computed numerically for the absorption of massless scalar waves, in the BH branch of interpolation of the line element \eqref{Lineel}. In order to compute numerically the absorption cross section, we integrate Eq.~\eqref{rad_eq}, using the boundary conditions~\eqref{BC_BH}. Then, we compute the transmission coefficient using the numerical solution for $\phi(x)$ and its derivative, and, hence, the absorption cross section is obtained through Eqs.~\eqref{tot_abs} and \eqref{partial_abs}. 

\begin{center}
\begin{figure}[ht!]
\center
\includegraphics[scale=0.68]{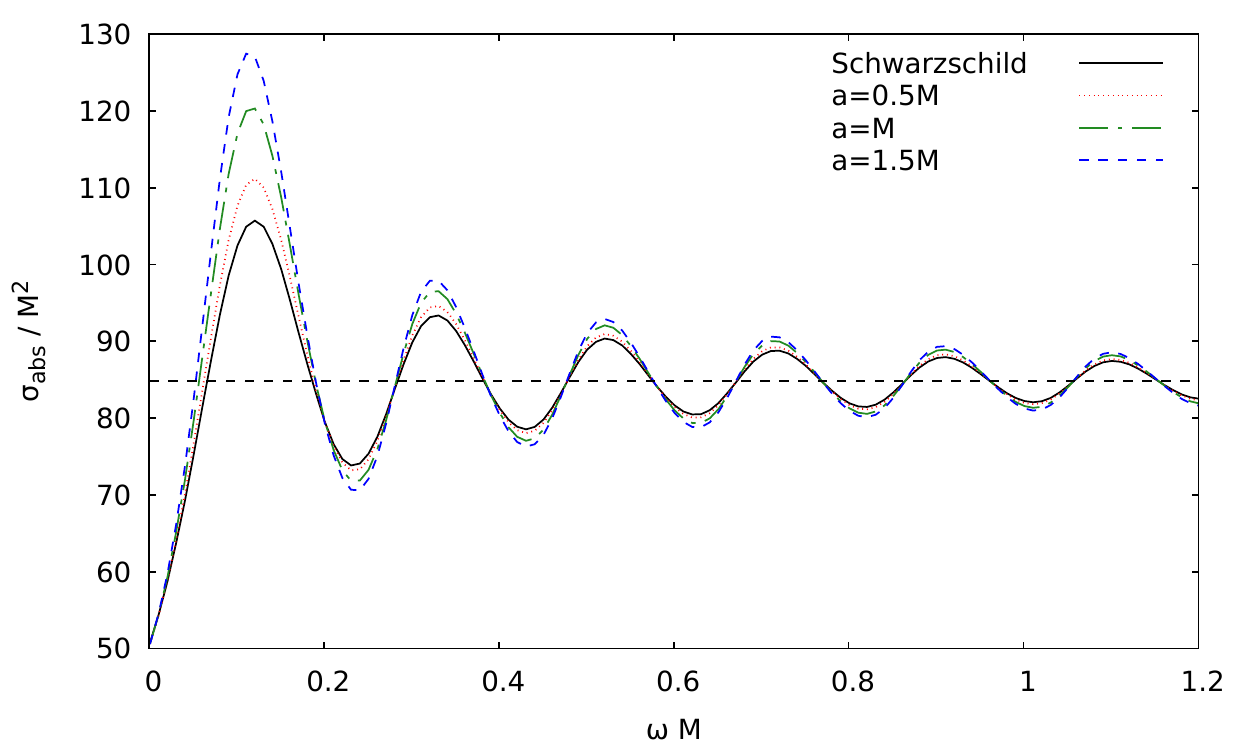}
\caption{Total absorption cross section of massless scalar waves for regular BHs (with different values of $a$), compared with the geometric cross section (horizontal dashed line). In this figure, we also show the total absorption cross section of massless scalar waves for the Schwarzschild BH, for comparison. }
\label{Total_abs}
\end{figure}
\end{center}

In Fig.~\ref{Total_abs}, we present the total absorption cross section of massless scalar waves for regular BHs with $a=0.5M, M,1.5M$, as well as the result for the Schwarzschild BH, for comparison purposes. The horizontal dashed line is the geometric cross section, which is the same for any value of $0\leq a \leq 3M $. We see in Fig.~\ref{Total_abs} that in the low-frequency regime, the total absorption cross section tends to the same value, regardless of the parameter $a$. This value is the area of the BH event horizon given in Eq.~\eqref{EH_area}.
In the high-frequency limit, the total absorption cross section tends to the same value, which is the geometric cross section for BHs, given in Eq.~\eqref{Geo_Cross_BH}. This result is in agreement with the discussion present in Sec.~\ref{HFR}. Furthermore, we see that as we increase the parameter $a$, the amplitude of oscillation of the total absorption cross section increases.

\begin{figure}[htp]
  \centering
  \subfigure[]{\includegraphics[scale=0.6]{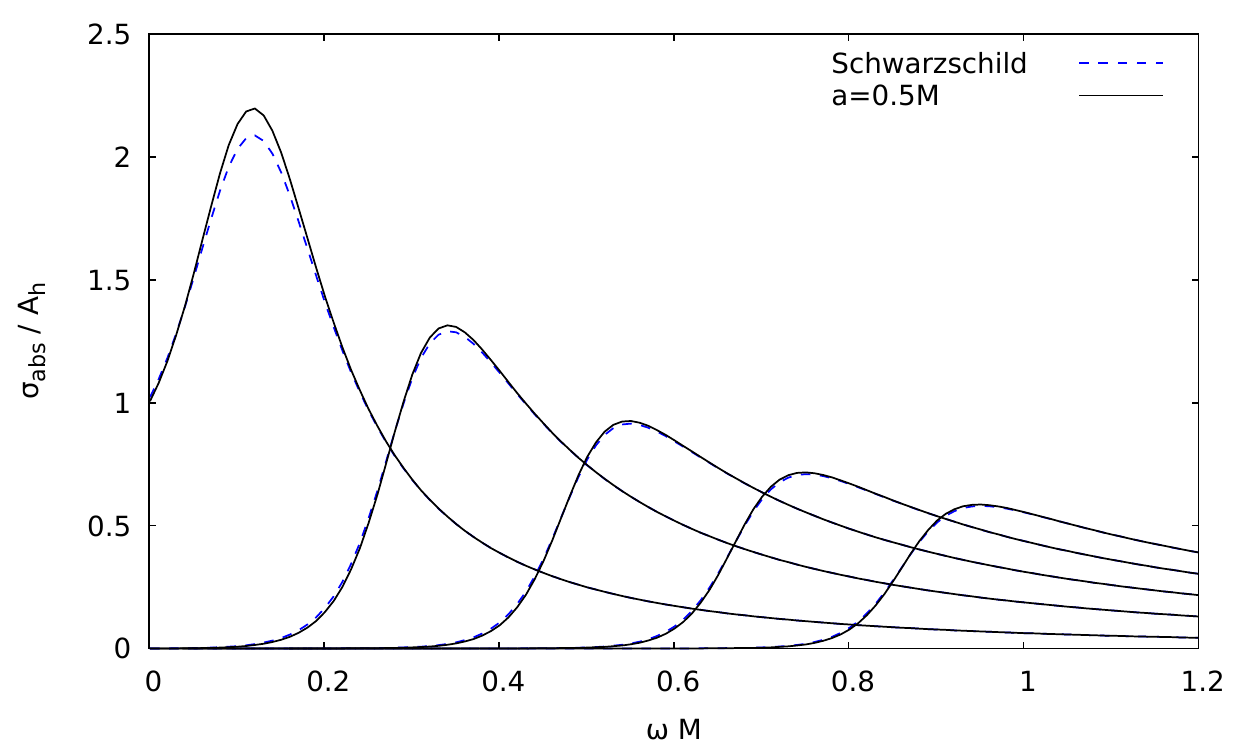}\label{a0}}\quad
  \subfigure[]{\includegraphics[scale=0.6]{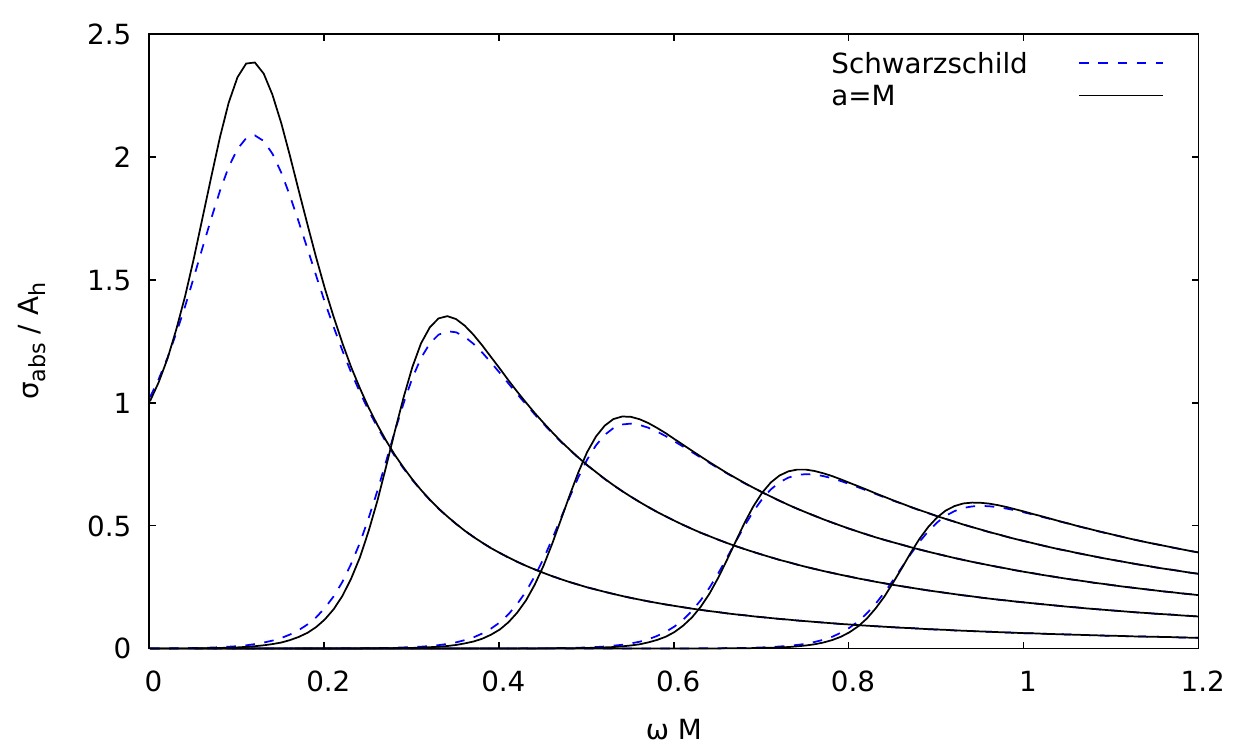}\label{b0}}
\subfigure[]{\includegraphics[scale=0.6]{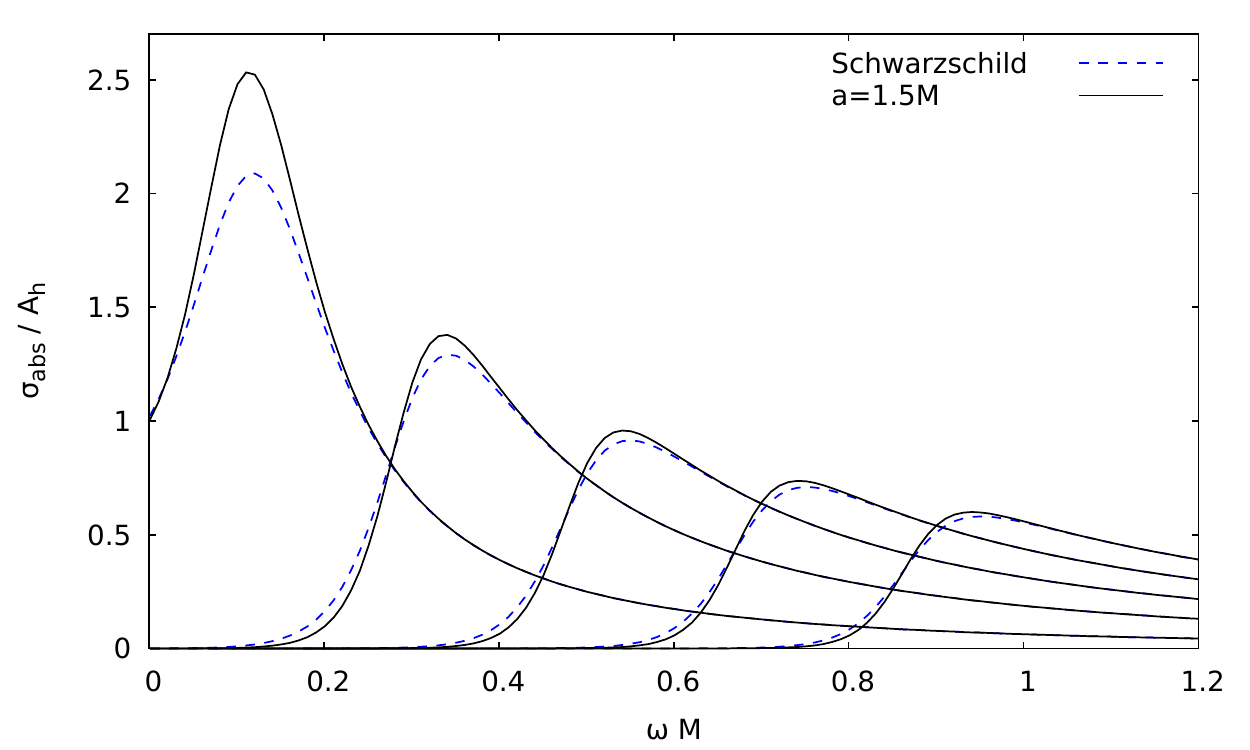}\label{c0}}
\caption{Partial absorption cross sections of massless scalar waves for regular BHs with different values of $a$. We also plot the partial absorption cross section for the Schwarzschild BH, for comparison. }\label{Fig.5}
\end{figure}

In Fig.~\ref{Fig.5}, we show the partial absorption cross sections in units of the event horizon area $A_h$, for regular BHs with $a=0.5M, M$, and $1.5M$, as well as for the Schwarzschild BH. We see that for $l=0$ and $\omega \rightarrow 0$, the partial absorption cross section $\sigma_0$ tends to the area of the BH, in agreement with the results presented in Sec.~\ref{Lowf_regime}.

\subsection{Discussion}

The results presented in this section show that the regular BHs described by Eq.~\eqref{Lineel} can mimic the Schwarzschild BH in the low- and high-frequency regimes of the absorption cross section, regardless of the parameter $a$. Results in the high-frequency regime play a representative role in the EHT observations, since they are directly related to BH shadows. Moreover, the regular BH absorption resembles the Schwarzschild BH case in the mid-to-high frequency regime, as can be seen in Fig.~\ref{Total_abs}. The main difference between the regular BHs and the Schwarzschild BH is manifest around the first local maximum of the absorption cross section. The increasing of the parameter $a$, implies in an increasing of the local maxima and a decreasing of the local minima of the total absorption cross section. 

\section{Wormhole results}
\label{Sec5}

\begin{center}
\begin{figure}
\includegraphics[scale=0.63]{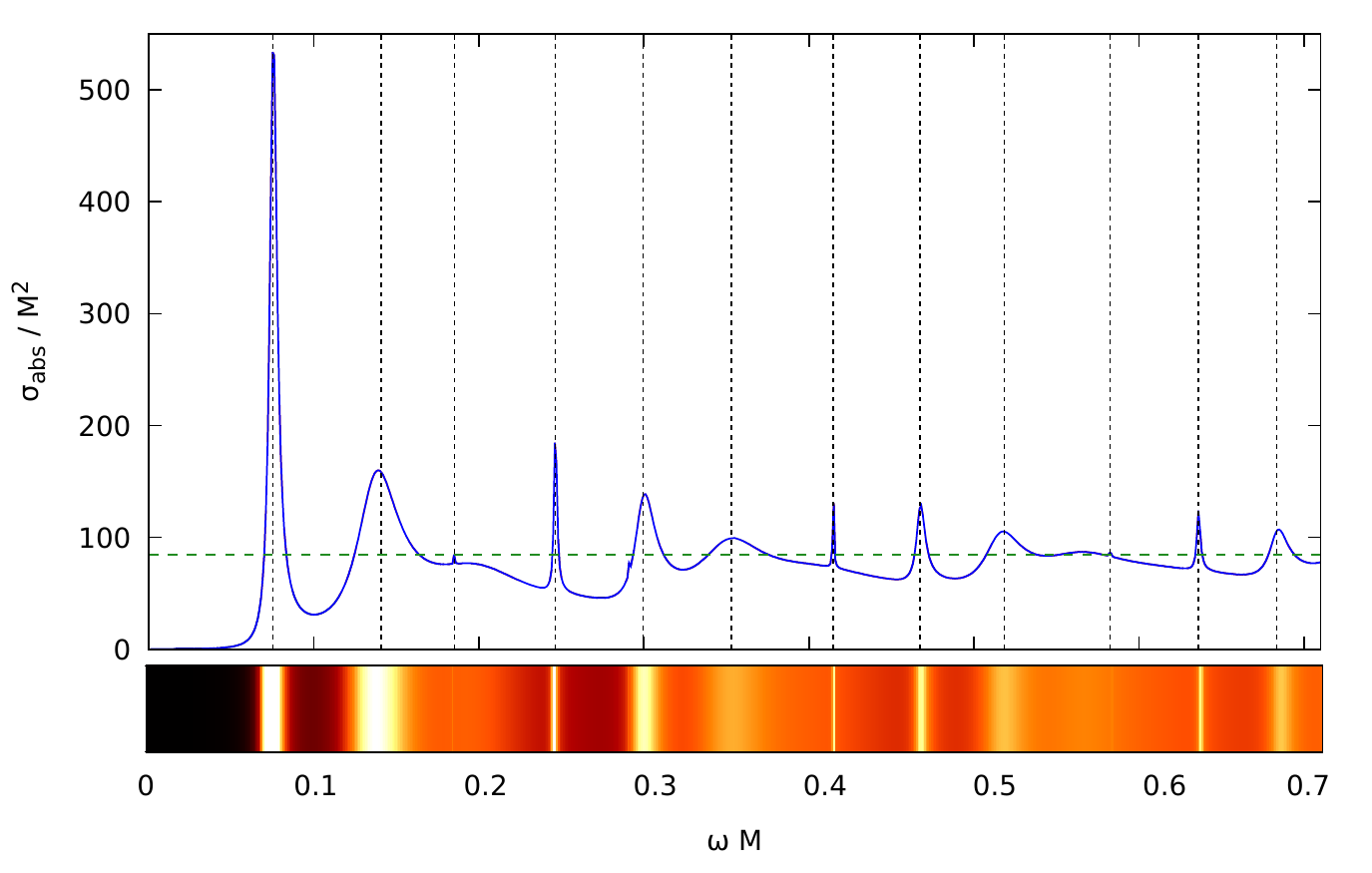}
\caption{Total absorption cross section of massless scalar waves for the wormhole with $a= 2.1M$, compared with the geometric cross section (horizontal line). The narrow peaks, associated to the vertical black dashed lines arise due to the potential well, which imply in the existence of the trapped modes. We also exhibit, below the plot, an absorption band composed with the results for the total absorption cross section.}
\label{Fig10}
\end{figure}
\end{center}

\begin{figure}[htp]
  \centering
  \subfigure[]{\includegraphics[scale=0.6]{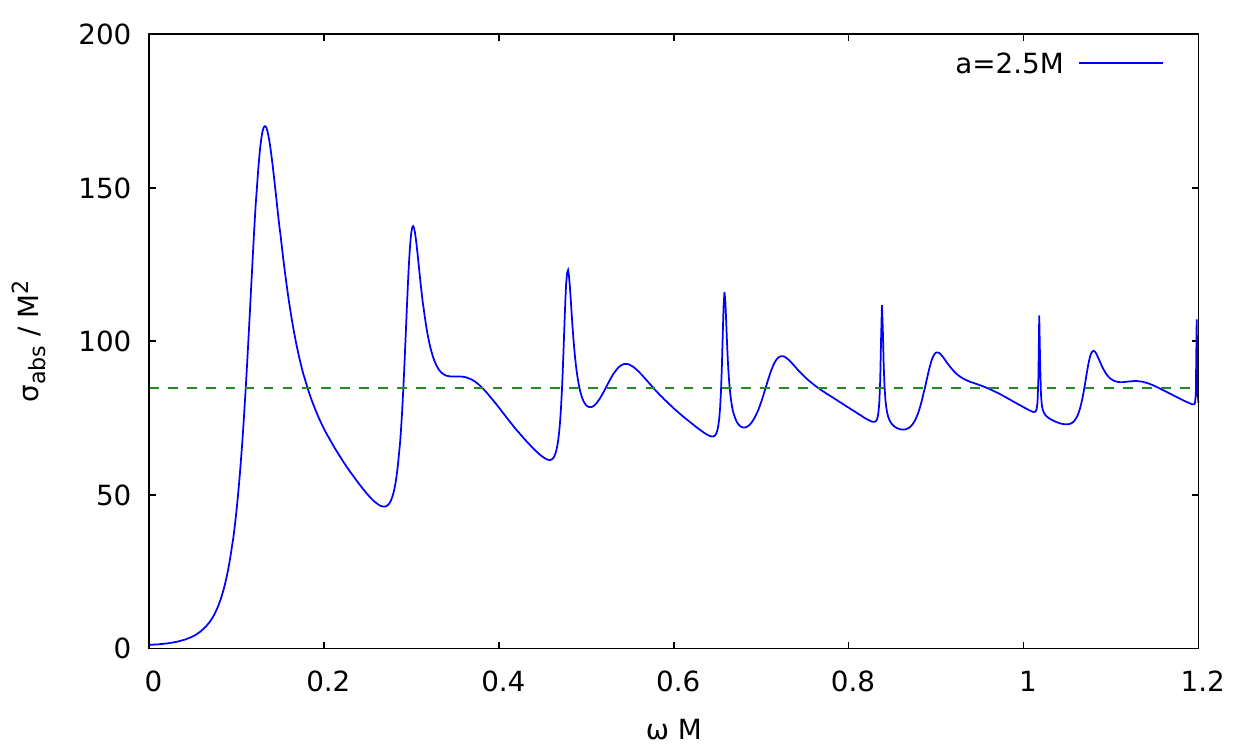}\label{b}}
\subfigure[]
{\includegraphics[scale=0.6]{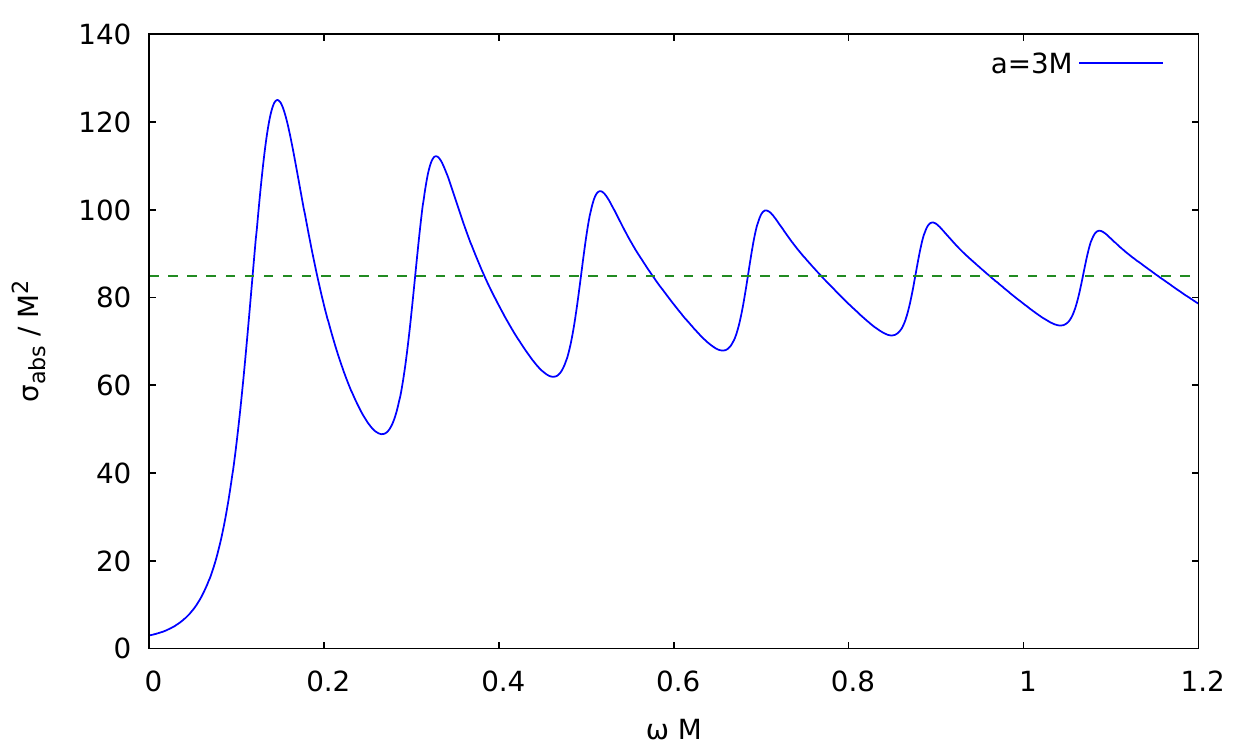}\label{c}}
\subfigure[]
{\includegraphics[scale=0.6]{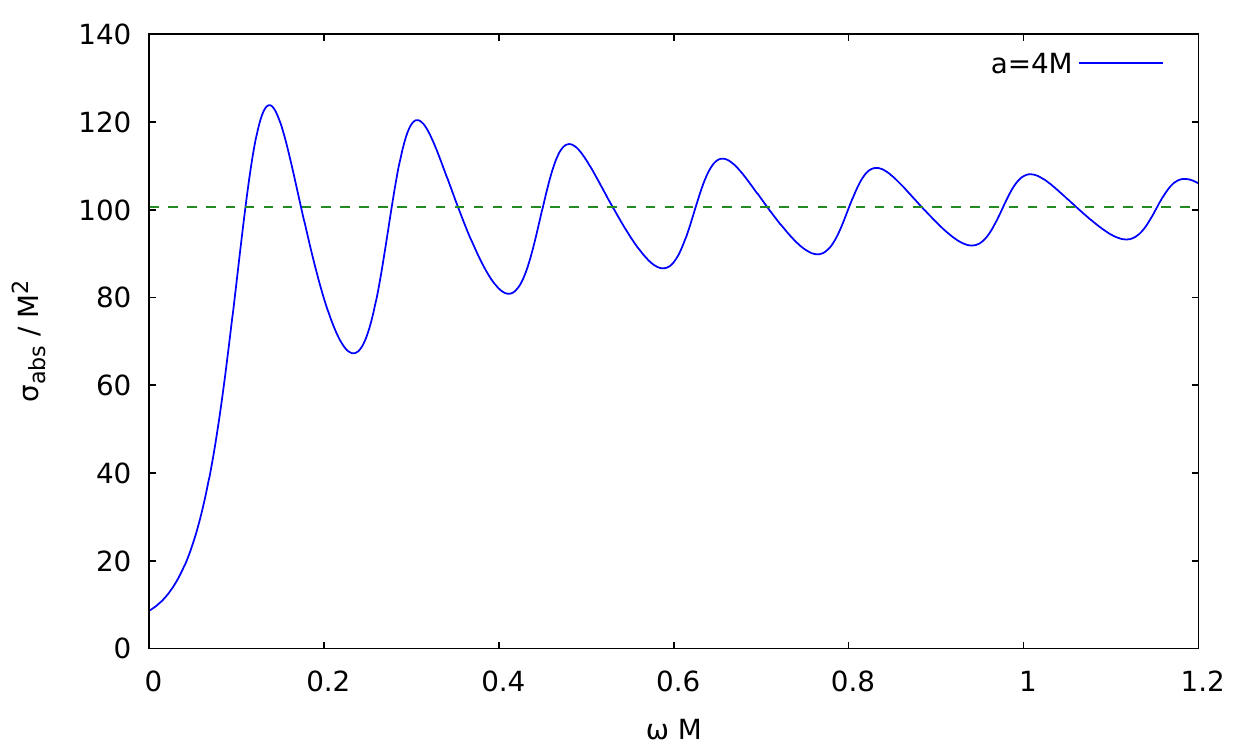}\label{d}}
\caption{Total absorption cross sections of massless scalar waves for wormholes with different values of $a$, compared with the geometric cross section (horizontal lines). In Figs.~\ref{b}, \ref{c}, and \ref{d}, we have set $a=2.5M$, $a=3M$ and $a=4M$, respectively. }\label{Fig.9}
\end{figure}

As an important feature of the Simpson-Visser line element, it may also represent a wormhole for $a\geq 2\,M$. In this section, we show the results computed numerically for the absorption cross section of massless scalar waves for the   wormhole branch of interpolation associated with the Simpson-Visser line element \eqref{Lineel}. 

Due to the shape of the potential of the wormhole case, the results are, in general, quite different from the BH ones. Such difference arises due to the presence of a potential well, as can be seen in Fig.~\ref{Veff2}. This potential well allows quasibound states to exist around $r=0$. These quasibound states are similar to the trapped modes discussed in Refs.~\cite{PROCRSOC, K&S,PRD:044069:2014,PRD:98104034:2018,PRD:100024016:2019, WMHL1,WMHL2}, and are associated to stable null geodesics at the wormhole throat in the eikonal limit. These trapped modes have complex frequency, and the imaginary part is usually small, i.~e. they are long-lived modes. 
They obey the following boundary conditions:
\begin{equation}
\label{QNM}\phi(r) = \begin{cases} e^{i\omega x},\ \ \ \ \ \  x\rightarrow \infty,\\
\,e^{-i\omega x}, \ \ \ \ x \rightarrow -\infty.
\end{cases}
\end{equation}
To find the frequencies associated to these trapped modes, we employ the numerical method proposed in Ref.~\cite{Direct_int}, known as direct integration method.\footnote{The computation of the complex frequencies of these trapped modes is analogous to what is done for quasi-normal mode (QNM) problems.}
By using the boundary conditions~\eqref{QNM}
in Eq.~\eqref{rad_eq}, we generate an eigenvalue problem for the frequency $\omega$. Frequencies that solve this eigenvalue problem are the trapped modes frequencies we are searching for. We numerically integrated Eq.~\eqref{rad_eq}, subjected to the boundary conditions \eqref{QNM}, and used a numerical root finding procedure to obtain the frequencies $\omega$ \cite{Direct_int, MPCG}. A selection of results for a representative case, $a=2.1M$, are presented in Table~\ref{QNMFREQ}, where $\omega_R$ and $\omega_I$ are the real and imaginary part of the trapped modes frequencies, respectively.

We show in Fig.~\ref{Fig10} the total absorption cross section for the wormhole with $a=2.1M$, where we can see the existence of narrow resonant peaks. 
In Fig.~\ref{Fig10}, we also plot, as vertical dashed lines, the real part of the trapped modes frequencies, presented in Table~\ref{QNMFREQ}. 
Hence, we see that the narrow peaks in the total absorption cross section are associated to the 
trapped modes around the wormhole's throat.

\begin{table}[htp]
\caption{Trapped modes frequencies for a=2.1M.}
\centering
\label{QNMFREQ}
\begin{ruledtabular}
\begin{tabular}{@{\hspace{0em}}  c @{\hspace{0em}} c @{\hspace{0em}} c  c c c}
\textit{l} &$\omega_{R}$ &  -i\,$\omega_I$& \textit{l} &$\omega_{R}$&-i\,$\omega_I$\\
\hline
0&\ \  0.0753\ \ \   & $2.673\,10^{-3}$ &1 & 0.1852  &$4.249\,10^{-5}$\\
0& \ \ \ 0.1408 \ \ \  & $8.304\,10^{-3}$ &1 & 0.2464 & $1.011\,10^{-3}$\\
 &  & & 1 &0.2996 & $7.578\,10^{-3}$\\
 &  & & 1 & 0.3532 & $9.417\,10^{-3}$\\

 2& 0.4147 & $3.432\,10^{-4}$  & 3& 0.6360 & $1.222\,10^{-3}$\\
 2 &  0.4674 & $3.154\,10^{-3}$ &  3 &  0.6834 & $6.861\,10^{-3}$\\
 2& 0.5184 & $9.281\,10^{-3}$ &  3& 0.5825 & $1.096\,10^{-4}$
\end{tabular}
\end{ruledtabular}
\end{table}

\begin{figure*}[ht!]
  \centering
  \subfigure[]{\includegraphics[scale=0.55]{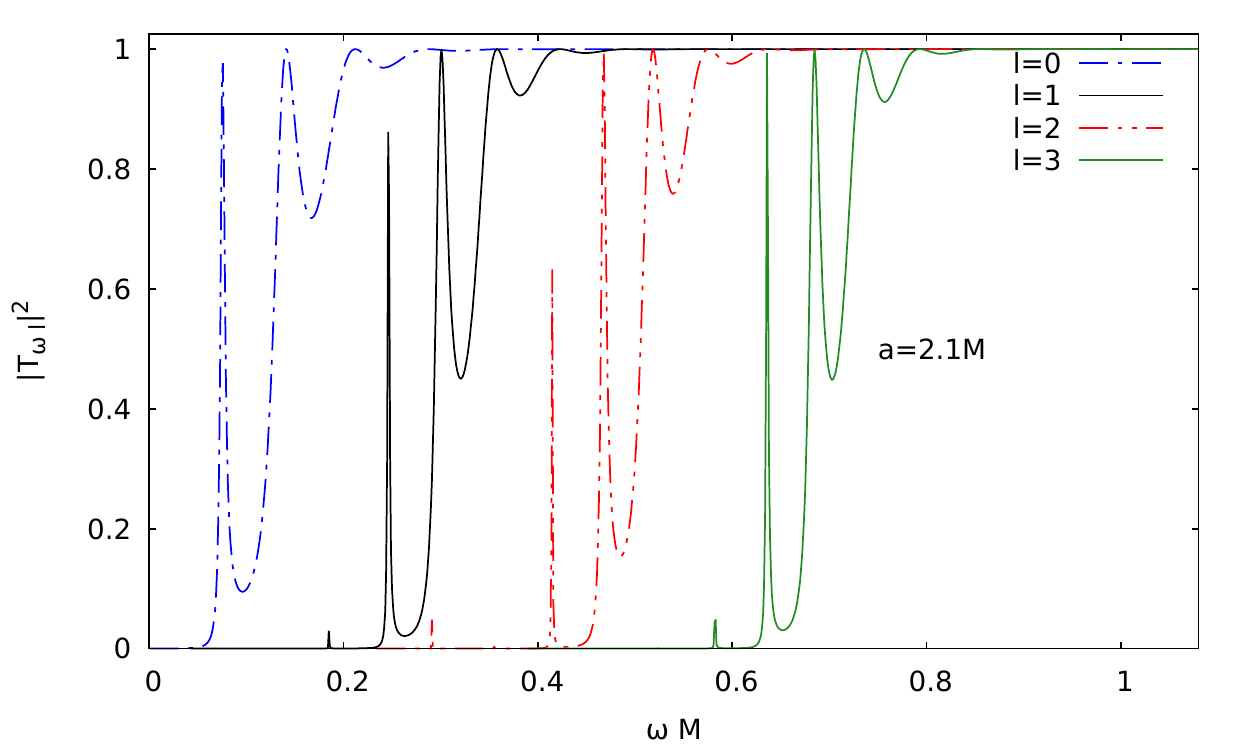}\label{a1}}
\subfigure[]
{\includegraphics[scale=0.55]{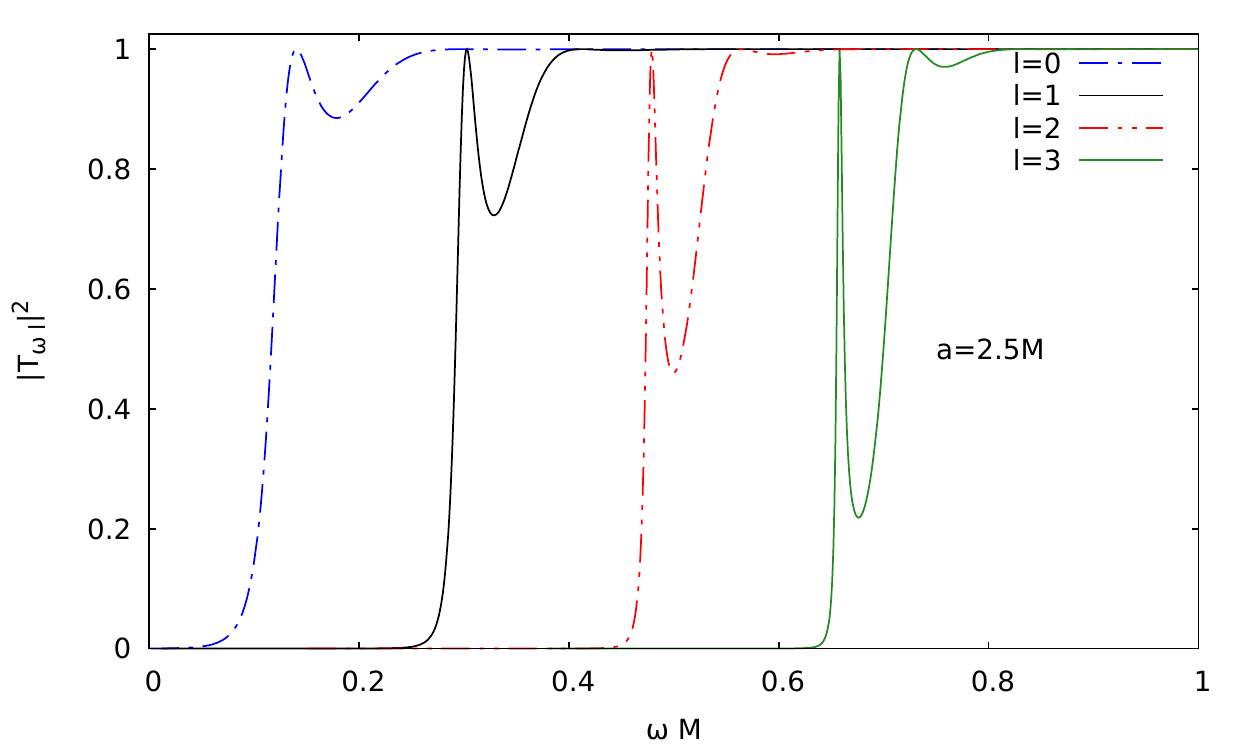}\label{b1}}
\subfigure[]
{\includegraphics[scale=0.55]{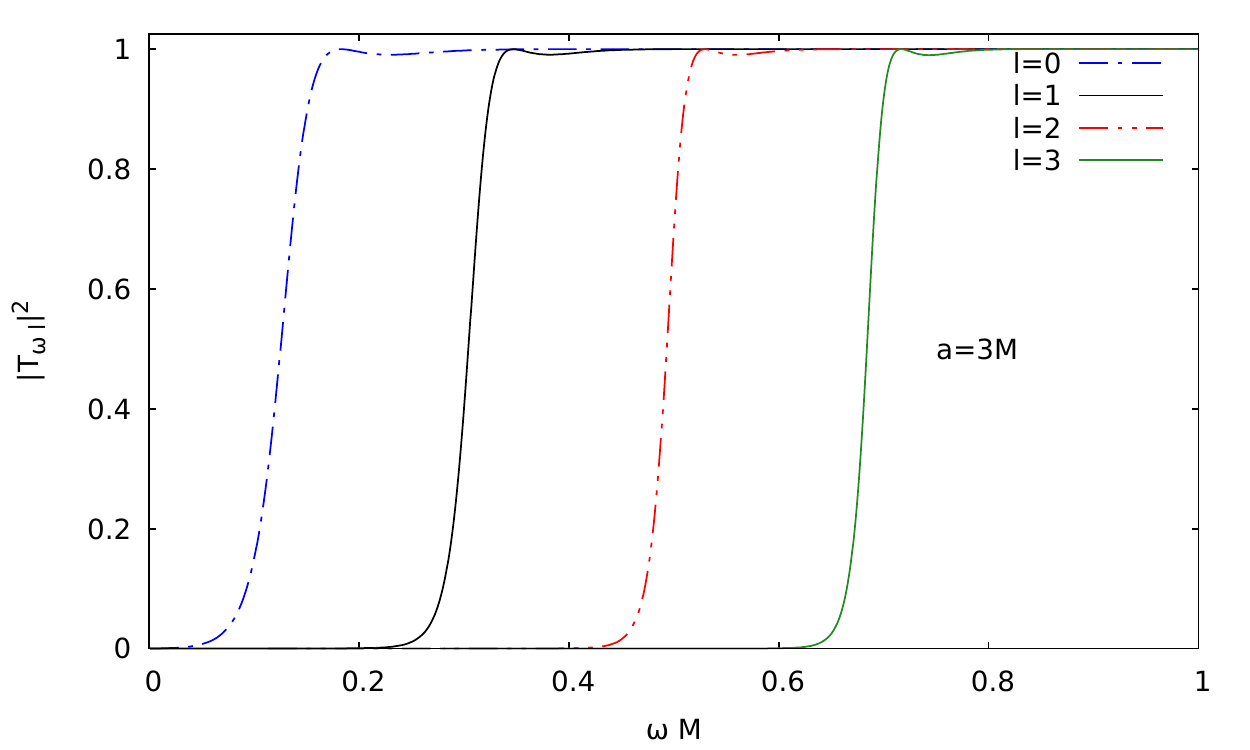}\label{c1}}
\subfigure[]
{\includegraphics[scale=0.55]{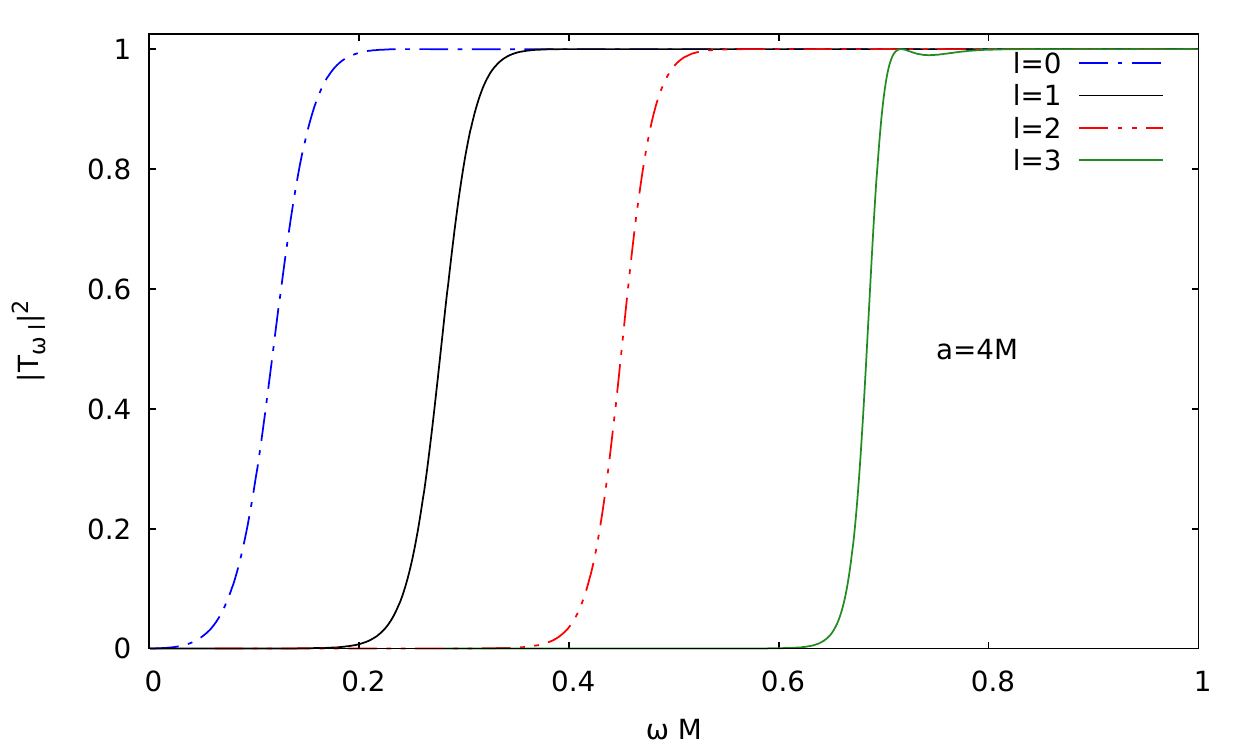}\label{d1}}
\caption{Transmission coefficients of massless scalar waves for wormholes with different values of $l$. In Figs.~\ref{a1}, \ref{b1}, \ref{c1},  and \ref{d1} we have set $a=2.1M$, $a=2.5M$, $a=3M$ and $a=4M$, respectively. We note that for $a=2.1M$, the peaks due to trapped modes can be easily identified. The peaks become less evident as we increase the value of $a$.}\label{Fig.11}
\end{figure*}

 In Fig.~\ref{Fig.9}, we present the total absorption cross section of massless scalar waves for wormholes with different values of the parameter $a$. We note from Fig.~\ref{Fig.9}, that as we increase the value of $a$, the local maxima become broader and the trapped modes peaks become less distinctive.  This is an effect of the decreasing in the potential well when $a$ increases (cf. Fig.~\ref{Veffwhml_geo}). We also note that the low-frequency limit of the total absorption cross section depends on the parameter $a$. From Fig.~\ref{d}, we note that the total absorption cross section behaves similarly to the BH case, presented in Fig.~\ref{Total_abs}.

We point out that for $a>3\,M$, the total absorption cross section, in the high-frequency limit, depends on the value of the parameter $a$ [see, for instance, Fig.~\ref{d}].   This result is in agreement with Eq.~\eqref{Geo_Cross_BH}.

In Fig.~\ref{Fig.11}, we present the results for the transmission coefficient, as a function of the frequency, for different values of $a$. We note that, as we increase the value of $a$, the number of peaks in the transmission coefficient  decreases. Using an approximation based on the Breit-Wigner expression for nuclei scattering, one can relate the amplitude of the transmission factor with the frequency of the trapped modes through \cite{PRD:98104034:2018,PR:49519:1936,PR:71145:1947}:
\begin{equation}
\label{BW_approx}|T_{\omega l}|^2=\frac{A_{\omega l}}{\left(\omega-\omega_R\right)^2+\omega_I^2},
\end{equation}
where $A_{\omega l}$ are constants, for a given $\omega$ and $l$, evaluated around each peak where $\omega\approx\omega_R$.  From Eq.~\eqref{BW_approx}, we can see that the peaks of the transmission coefficient happen at $\omega\approx\omega_R$, and the form of the peaks is related to $\omega_I$. 

In Fig.~\ref{Fig.12}, we plot the partial absorption cross sections of massless scalar waves for wormholes, with different choices of the parameter $a$. The partial absorption cross sections are related to the transmission coefficient, as given in Eq.~\eqref{partial_abs}. 

\subsection{Discussion}
\begin{figure*}[htp]
  \centering
  \subfigure[]{\includegraphics[scale=0.65]{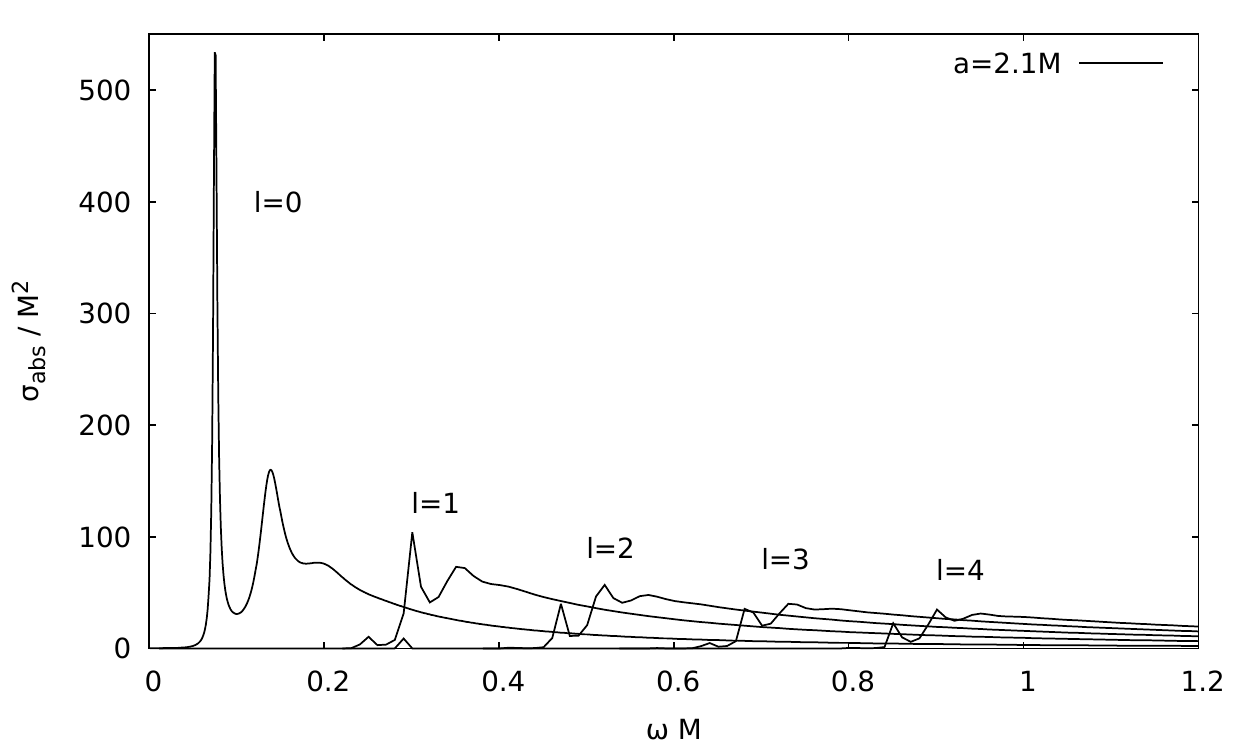}\label{a2}}\quad
  \subfigure[]{\includegraphics[scale=0.65]{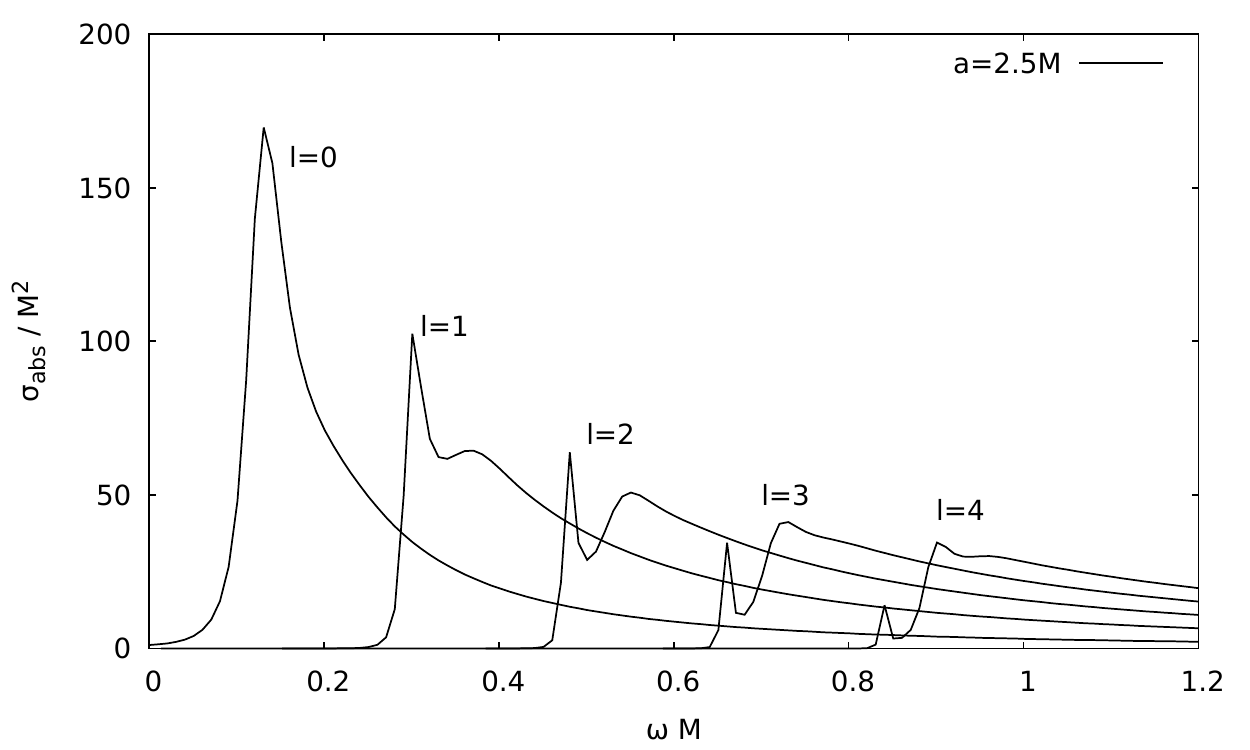}\label{b2}}
\subfigure[]
{\includegraphics[scale=0.65]{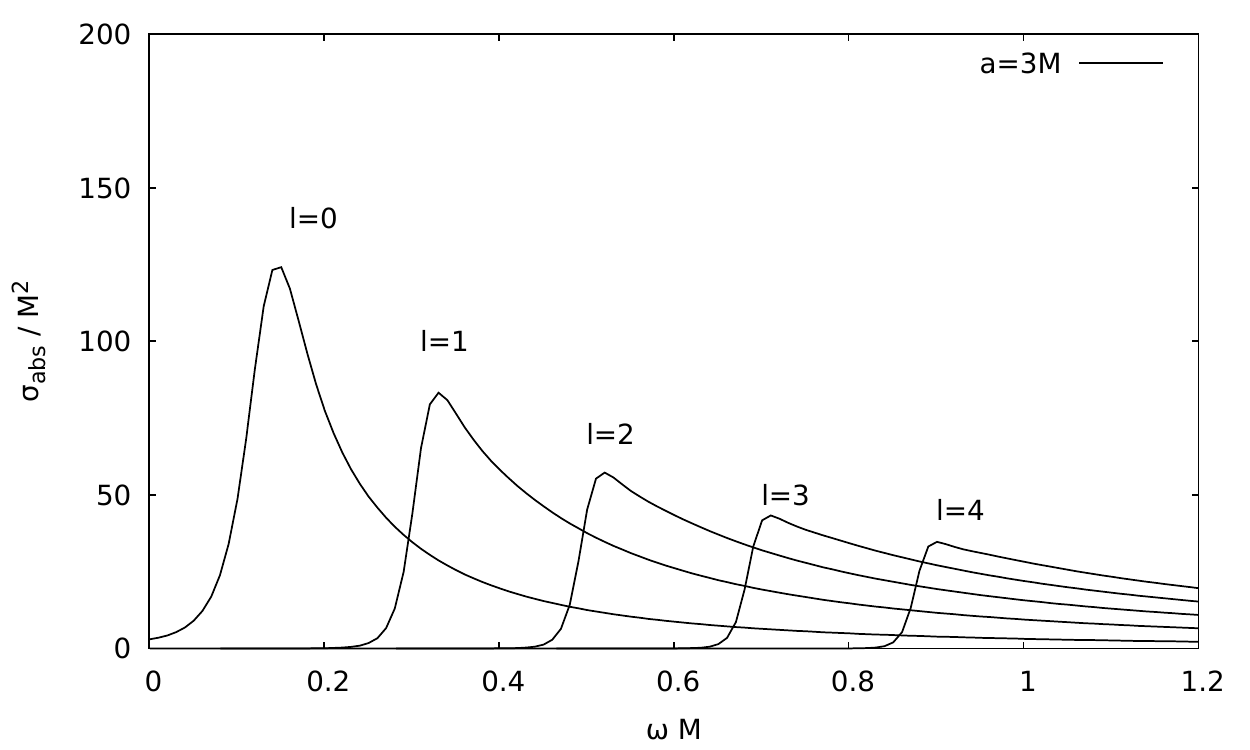}\label{c2}}
\subfigure[]
{\includegraphics[scale=0.65]{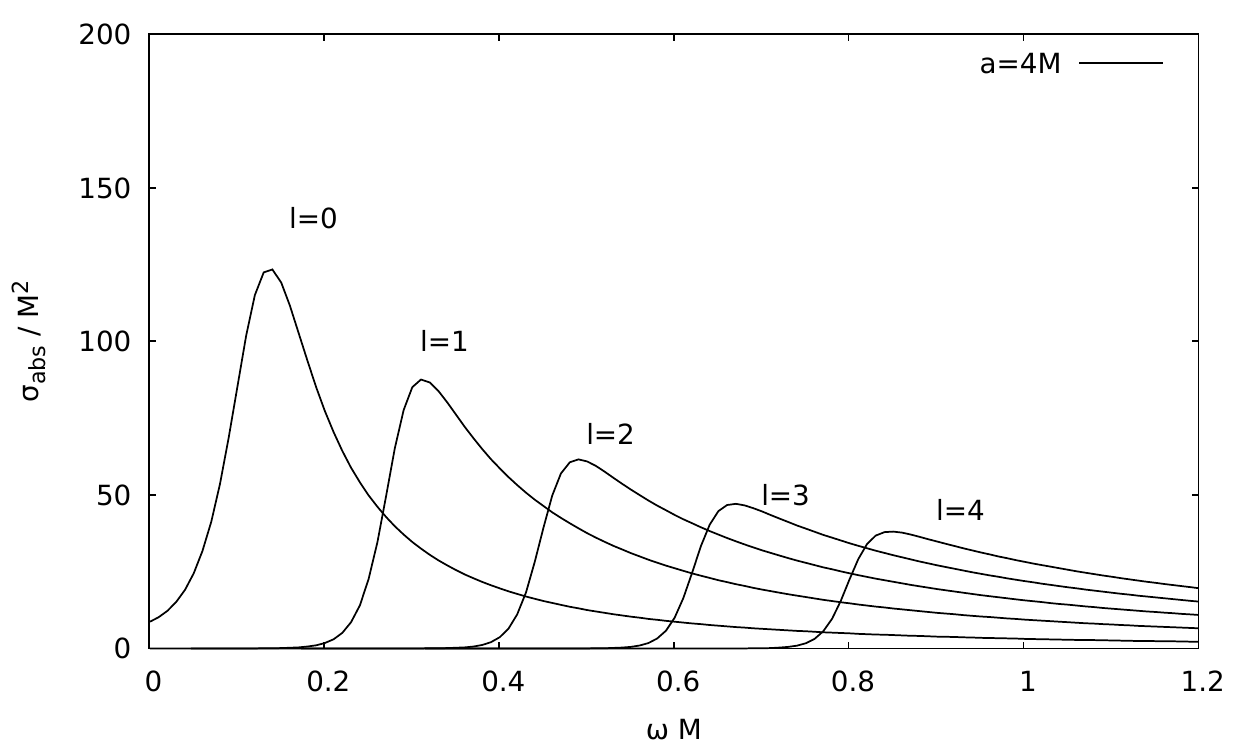}\label{d2}}
\caption{Partial absorption cross sections of massless scalar waves for the wormhole, with different values of $a$. In Figs.~\ref{a2}, \ref{b2}, \ref{c2}, and \ref{d2} we have set $a=2.1M$, $a=2.5M$, $a=3M$ and $a=4M$, respectively. As in Fig.~\ref{Fig.11}, for $a=2.1M$ and $a=2.5M$, we see that the peaks due to trapped modes are easy to identify. The peaks become less evident as we increase the value of $a$.}\label{Fig.12}
\end{figure*}

From Fig.~\ref{Fig.12}, we see that the resonant peaks become broader and 
less distinct as the parameter $a$ increases, what is consistent with the behavior of the potential shown in Fig.~\ref{Veff2}. These resonances can make the wormhole absorption spectrum  quite distinctive from the BH one. Similar results are also manifest in extreme compact objects (ECOs)~\cite{PRD:98104034:2018} and in BH remnants~\cite{PRD:100024016:2019} emerging on quadratic and Born-Infeld theories of gravity.  
Moreover, we note that the trapped modes, associated to the resonant peaks,  are slowly decaying modes, since the imaginary part is always negative and small, as can be seen in Table~\ref{QNMFREQ}. It is worth mentioning that trapped modes in wormhole spacetimes also play a role in gravitational waveforms, giving rise to echoes~(see, e. g., Ref.~\cite{WMHL2,Cardoso:2016rao}).

\section{Final Remarks}
\label{Sec6}
We have studied the propagation of massless scalar waves in the spacetime configuration proposed by Simpson and Visser~\cite{Black_bounce}.
This Simpson-Visser geometry is characterized by a parameter that allows the interpolation from BH to wormhole configurations.

We have investigated the scalar absorption by BHs and wormholes associated to the Simpson-Visser configuration. For the BH branch of interpolation, we have shown that the low- and high-frequency results are equivalent to the Schwarzschild BH ones, namely, they are equal to the BH horizon area and tend to the geometric capture cross section, respectively. Moreover, the generic oscillation behavior around the high-frequency limit of spherical BHs is manifest. Our numerical results are in full agreement with the low- and high-frequency regime approximations, and also with the sinc approximation.

We found that the BH and the wormhole configurations can be quite distinctive concerning the absorption of scalar waves.
The distinction is due to the presence of trapped modes around the wormhole's throat and to the different values of the total absorption cross section in the low- and high-frequency limits.
The absorption cross section of the wormhole branch of interpolation can present narrow resonant peaks due to a potential well at the throat of the wormhole. 
Moreover, these peaks become broader as we increase the parameter $a$, due to the decreasing of the depth of the potential well around the wormhole throat. 
Similar resonance effects in the absorption cross section were reported for ECOs~\citep{PRD:98104034:2018} and for BH remnants~\cite{PRD:100024016:2019}, where the partial transmission amplitudes also present Breit-Wigner-type resonances, analogously to the phenomenon present in nuclear scattering theory \cite{PR:49519:1936,PR:71145:1947}.

\acknowledgments
The authors thank 
Conselho Nacional de Desenvolvimento Cient\'ifico e Tecnol\'ogico (CNPq) and Coordena\c{c}\~ao de Aperfei\c{c}oamento de Pessoal de N\'{\i}vel Superior (Capes) - Finance Code 001, for partial financial support. We are grateful to Carlos Herdeiro and Pedro Cunha for useful discussions.
This research has also received funding from the European Union's Horizon 2020 research and innovation programme under the H2020-MSCA-RISE-2017 Grant No. FunFiCO-777740.


\begin{thebibliography}{9}
\bibitem{VL1} B.~P.~Abbott, et al., Observation of gravitational waves from a binary black hole merger. Phys. Rev. Lett. \textbf{116}, 061102 (2016).

\bibitem{VL2} B.~P.~Abbott, et al., GW151226: observation of gravitational waves from a 22-solar-mass binary black hole coalescence. Phys. Rev. Lett. \textbf{116}, 241103 (2016).

\bibitem{VL3} B.~P.~Abbott, et al., GW170104: observation of a 50-solar-mass binary black hole coalescence at redshift 0.2. Phys. Rev. Lett. \textbf{188}, 221101 (2017). 

\bibitem{VL4} B.~P.~Abbott, et al., GW170608: observation of a 19 solar-mass binary black hole coalescence, Astrophys. J. Lett. \textbf{851}, L35 (2017).

\bibitem{EHT} The Event Horizon Telescope Collaboration, First M87
event horizon telescope results. I. The shadow of the
supermassive black hole, Astrophys. J. Lett. \textbf{875}, L1
(2019).

\bibitem{Sch} K.~Schwarzschild, Sitzungsber.~Preuss.~Akad.~Wiss.~Berlin (Math.~Phys.) \textbf{1916}, 189 (1916).

\bibitem{Reissner}H.~Reissner, {\"U}ber die Eigengravitation des elektrischen Feldes nach der Einsteinschen theorie. Annalen der Physik \textbf{355}, 106 (1916).


\bibitem{Nordstrom}G.~Nordstr{\"o}m, On the energy of gravitation field in Einstein's theory. Kon.~Ned.~Akad.~Wet., \textbf{20}, 1238 (1918).

\bibitem{Kerr:1963} R.~P.~Kerr, Gravitational Field of a Spinning Mass as an Example of Algebraically Special Metrics. Phys.\ Rev.\ Lett.\ {\bf 11}, 237 (1963).

\bibitem{kerrnewman} E.~T.~Newman, E.~Couch, K.~Chinnapared, A.~Exton, A.~Prakash and R.~Torrence, Metric of a rotating, charged mass. J.\ Math.\ Phys. {\bf 6}, 918 (1965).

\bibitem{Penrose:1964wq} 
  R.~Penrose,
  Gravitational collapse and space-time singularities.
  Phys.\ Rev.\ Lett.\  {\bf 14}, 57 (1965).

\bibitem{Singularitytheo} S.~W.~Hawking and G.~F.~R.~ Ellis, \textit{The Large Scale Structure of Space-Time} (Cambridge University Press, Cambridge, 1973).

\bibitem{bardeen} J.~M.~Bardeen, Non-singular general-relativistic gravitational collapse. Proceedings of the International Conference GR5, 174 (1968).

\bibitem{Hayward} S.~A.~Hayward, Formation and evaporation of nonsingular black holes. Phys.\ Rev.\ Lett.\ {\bf 96},  031103 (2006).

\bibitem{ayonb} E.~A.~Beato and A.~Garc\'{\i}a, New regular black hole solution from nonlinear electrodynamics. Phys.\ Lett.\ B {\bf 464}, 25 (1999).

\bibitem{RBH} V.~P.~Frolov, Notes on nonsingular models of black holes. Phys.~Rev.~D \textbf{94}, 104056 (2016).

\bibitem{Matt_Visser} M.~Visser, \textit{Lorentzian wormholes: from Einstein to Hawking} (American Institute of Physics, New York, 1995).

\bibitem{Einstein_Rosen}A.~Einstein and N.~Rosen, The particle problem in the general theory of relativity. Phys.~Rev.~\textbf{48}, 73 (1935).

\bibitem{dinverno} R.~D'Inverno,  \textit{Introducing Einstein's Relativity} (Clarendon Press,
Oxford, 1992).
 

\bibitem{MorrisThorne} M.~S.~Morris and K.~S.~Thorne, Wormholes in spacetime and their use for interstellar travel: A tool for teaching general relativity. Am. J. Phys. \textbf{56}, 395 (1988).

\bibitem{Morris-Thorne-Yurtsever} M.~S.~Morris, K.~S.~Thorne and U.~Yurtsever, Wormholes, Time machines, and the weak energy condition. Phys.~Rev.~Lett.~ \textbf{61}, 1446 (1988).

\bibitem{Matt_Visser::1989} M.~Visser, Traversable wormholes: Some simple examples. Phys.~Rev.~D~\textbf{39}, 3181(R) (1989)

\bibitem{AGN} G.~L.~Granato, G.~D.~Zotti, L. Silva, A.~Bressan and L.~Danese, A physical model for the coevolution of QSOs and their spheroidal hosts. Astrophys. J. \textbf{600}, 580 (2004).

\bibitem{AGN1}A.~Marconi, G.~Risaliti, R.~Gilli, L.~K.~Hunt, R.~Maiolino and M.~Salvati, Local supermassive black holes, relics of active galactic nuclei and the X-ray background. Mon. Not. R. Astron. Soc. \textbf{351}, 169 (2004).

\bibitem{AGN2} L.~Ferrarese and H.~Ford, Supermassive black holes in galactic nuclei: Past, present and future research. Space Sci. Rev. \textbf{116}, 523 (2005).

\bibitem{Abs1}N. Sanchez, Absorption and emission spectra of a Schwarzschild black hole. Phys. Rev. D \textbf{18}, 1030 (1978).


\bibitem{Abs2} T.~Nakamura and H.~Sato, Absorption of massive scalar field by a charged black hole. Phys.~Lett.~B \textbf{61}, 371 (1976).

\bibitem{Abs4} C.~L.~Benone, E.~S.~Oliveira, S.~R.~Dolan and L.~C.~B.~Crispino, Absorption of a massive scalar field by a charged black hole. Phys.~Rev.~D \textbf{89}, 104053 (2014).

\bibitem{Abs5} C.~F.~B.~Macedo and L.~C.~B.~Crispino, Absorption of planar massless scalar waves by Bardeen black holes. Phys.~Rev.~D \textbf{90}, 064001 (2014).

\bibitem{Abs6} C.~F.~B.~Macedo, L.~C.~S.~Leite, E.~S.~Oliveira, S.~R.~Dolan and L.~C.~B.~Crispino, Absorption of planar massless scalar waves by Kerr black holes. Phys.~Rev.~D \textbf{88}, 064033 (2013).

\bibitem{Abs8} L.~C.~S.~Leite, C.~L.~Benone and L.~C.~B.~Crispino, Scalar absorption by charged rotating black holes. Phys.\ Rev.\ D \textbf{96}, 044043 (2017).

\bibitem{Abs9} C.~L.~Benone, L.~C.~S.~Leite, L.~C.~B.~Crispino and 
S.~R.~Dolan, On-axis scalar absorption cross section of Kerr-Newman black holes: Geodesic analysis, sinc and low-frequency approximations. Int.\ J.\ Mod.\ Phys.\ D \textbf{27}, 1843012 (2018).

\bibitem{Abs10} C.~L.~Benone and L.~C.~B.~Crispino, Massive and charged scalar field in Kerr-Newman spacetime: Absorption and superradiance. Phys.\ Rev.\ D\ \textbf{99}, 044009 (2019).

\bibitem{Abs11} H.~Huang, J.~Chen, Y.~Wang and Y.~Jin, Absorption of a massive scalar field by Wormhole space-times. Int.~J.~Theor.~Phys.~ \textbf{56}, 1150 (2017).

\bibitem{PRD:100024016:2019} A.~Delhom, C.~F.~B.~Macedo, G.~J.~Olmo and L.~C.~B.~Crispino, Absorption by black hole remnants in metric-affine gravity. Phys.~Rev.~D \textbf{100}, 024016 (2019). 

\bibitem{Black_bounce} A. Simpson and M. Visser, Black-Bounce to traversable wormhole. J. Cosmol. Astro. Phys. JCAPP02(2019)042. 

\bibitem{QNM_SV} M.~S.~Churilova and Z.~Stuchl{\'i}k, Ringing of the regular black-hole/wormhole transition. Class.~Quantum~Grav.~\textbf{37}, 075014 (2020).

\bibitem{Black_bounce2}H.~Huang and J.~Yang, Charged Ellis wormhole and black bounce. Phys.~Rev.~D \textbf{100}, 124063 (2019).

\bibitem{PRD:98104034:2018} C.~F.~B.~Macedo, T.~Stratton, S.~R.~Dolan and L.~C.~B.~Crispino, Spectral lines of extreme compact objects. Phys.~Rev.~D \textbf{98}, 104034 (2018).

\bibitem{Futterman} J.~A.~H.~Futterman, F.~A.~Handler and R.~A.~Matzner, \textit{Scattering from Black Holes} (Cambridge University Press, Cambridge, 1988).

\bibitem{Wald} R.~M.~Wald, \textit{General Relativity} (The University of Chicago Press, Chicago, 1984).

\bibitem{Das_Gibbons_Mathur} S.~R.~Das, G.~W.~Gibbons and S.~D.~Mathur, Universality of low energy absorption cross sections for black holes. Phys. Rev. Lett. \textbf{78}, 417 (1997).

\bibitem{AH1} A. Higuchi, Low-frequency scalar absorption cross sections for stationary black holes. Class. Quantum Grav. \textbf{18}, L139 (2001); Addendum, Class. Quantum Grav. \textbf{19}, 599(A) (2002). 

\bibitem{Sinc_approx} Y. Decanini, G. Esposito-Farese, and A. Folacci, Universality of high-energy absorption cross sections for black holes. Phys. Rev. D \textbf{83}, 044032 (2011). 


\bibitem{Sinc_approx2} V. Cardoso, A. S. Miranda, E. Berti, H. Witek, and V.~T. Zanchin, Geodesic stability, Lyapunov exponents, and quasinormal modes. Phys. Rev. D \textbf{79}, 064016 (2009).


\bibitem{PROCRSOC} S.~Chandrasekhar and V.~Ferrari, On the non-radial oscillations of a star. III. A reconsideration of the axial modes. Proc.~R.~Soc.~A \textbf{434}, 449 (1991).

\bibitem{K&S} K.~D.~Kokkotas and B.~G.~Schmidt, Quasinormal modes of stars and black holes. Living Rev. Relativity \textbf{2}, 2 (1999).

\bibitem{PRD:044069:2014} V.~Cardoso, L.~C.~B.~Crispino, C.~F.~B.~Macedo, H.~Okawa and P.~Pani, Light rings as observational evidence for event horizons: Long-lived modes, ergoregions and nonlinear instabilities of ultracompact objects. Phys.~Rev.~D \textbf{90}, 044069 (2014).

\bibitem{WMHL1} T.~Damour and S.~N.~Solodukhin, Wormholes as black holes foils. Phys.~Rev.~D~\textbf{76}, 024016 (2007).

\bibitem{WMHL2} P.~Bueno, P.~A.~Cano, F.~Goelen, T.~Hertog and B.~Vercnocke, Echoes of Kerr-like wormholes. Phys.~Rev.~D \textbf{97}, 024040 (2018).


\bibitem{Cardoso:2016rao} V.~Cardoso, E.~Franzin and P.~Pani, Is the gravitational-wave ringdown a probe of the event horizon?. Phys. Rev. Lett. \textbf{116}, 171101 (2016) [Erratum: Phys. Rev. Lett. 117 (2016) 8, 089902].


\bibitem{Direct_int} S.~Chandrasekhar and S.~Detweiler, The quasi-normal modes of the Schwarzschild black hole. Proc.~R.~Soc.~Lond.~A. \textbf{344}, 441~(1975).

\bibitem{MPCG} C. Molina, P. Pani, V. Cardoso, and L. Gualtieri, Gravitational signature of Schwarzschild black holes in dynamical Chern-Simons gravity. Phys. Rev. D {\bf 81}, 124021 (2010).

\bibitem{PR:49519:1936} G.~Breit and E.~Wigner, Capture of slow neutrons. Phys.~Rev.~\textbf{49}, 519 (1936). 

\bibitem{PR:71145:1947} H.~Feshbach, D.~C.~Peaslee and V.~F.~Weisskopf, On the scattering and absorption of particles by atomic nuclei. Phys.~Rev.~\textbf{71}, 145 (1947).



\end{thebibliography}
\end{document}